\documentclass[aps,twocolumn,superscriptaddress,preprintnumbers,prl]{revtex4}
\usepackage{xr-hyper}
\usepackage{mathrsfs, hyperref}
\usepackage{amssymb, amsbsy, amsmath, latexsym, dsfont, array, layout, graphics,mathrsfs,braket,amsfonts,amsthm,
  amssymb,graphicx,subfigure, youngtab,color,bm,mathtools,braket,verbatim,url,cleveref,natbib,hypernat,extarrows,inputenc,multirow,bbm}
  
\usepackage[usernames,dvipsnames]{xcolor}
\colorlet{RED}{red}
\usepackage{graphicx,psfrag}
\usepackage{amsfonts}
\usepackage[figuresright]{rotating}  
\usepackage{amssymb}
\usepackage{amsmath}
\usepackage{subfigure}
\usepackage{multirow}
\usepackage{tabularx}
\usepackage[resetlabels]{multibib}
\newcites{supp}{Supplemental References}
\usepackage{amssymb, amsbsy, amsmath, latexsym, dsfont, array, layout, graphics,mathrsfs,braket,amsfonts,amsthm,
  amssymb,graphicx,subfigure,dcolumn, youngtab,color,bm,mathtools,braket,verbatim,url,cleveref,natbib,hypernat,extarrows,inputenc,multirow}
\usepackage[usernames,dvipsnames]{xcolor}
\graphicspath{ {./images/} }

\usepackage{mathrsfs, hyperref}
\usepackage{amssymb, amsbsy, amsmath, latexsym, dsfont, array, layout, graphics,mathrsfs,braket,amsfonts,amsthm,
  amssymb,graphicx,subfigure, youngtab,color,bm,mathtools,braket,verbatim,url,cleveref,natbib,hypernat,extarrows,inputenc,multirow,bbm}
\usepackage{graphicx}
\usepackage{dcolumn}
\usepackage{bm}
\usepackage[usernames,dvipsnames]{xcolor}
\usepackage[normalem]{ulem}
\usepackage[T1]{fontenc}
\graphicspath{ {./images/} }

\begin{document}
\title{Topological exact flat bands in two dimensional materials under periodic strain}

\author{Xiaohan Wan}
\affiliation{%
Department of Physics, University of Michigan, Ann Arbor, MI 48109, USA
}
\affiliation{%
Theoretical Division, T-4 and CNLS, Los Alamos National Laboratory, Los Alamos, New Mexico 87545, USA
}
\author{Siddhartha Sarkar}
\affiliation{%
Department of Physics, University of Michigan, Ann Arbor, MI 48109, USA
}
\author{Shi-Zeng Lin}
\email{szl@lanl.gov}
\affiliation{%
Theoretical Division, T-4 and CNLS, Los Alamos National Laboratory, Los Alamos, New Mexico 87545, USA
}
\affiliation{%
Center for Integrated Nanotechnologies (CINT), Los Alamos National Laboratory, Los Alamos, New Mexico 87545, USA
}
\author{Kai Sun}
\email{sunkai@umich.edu}
\affiliation{%
Department of Physics, University of Michigan, Ann Arbor, MI 48109, USA
}

\begin{abstract}
We study flat bands and their topology in 2D materials with quadratic band crossing points (QBCPs) under periodic strain. In contrast to Dirac points in graphene, where strain acts as a vector potential, strain for QBCPs serves as a director potential with angular momentum $\ell=2$. 
We prove that when the strengths of the strain fields hit certain ``magic" values, exact flat bands with $C=\pm 1$ emerge at charge neutrality point in the chiral limit, in strong analogy to magic angle twisted bilayer graphene. These flat bands have ideal quantum geometry for the realization of fractional Chern insulators, and they are always fragile topological. The number of flat bands can be doubled for certain point group, and the interacting Hamiltonian is exactly solvable at integer fillings.
We further demonstrate the stability of these flat bands against deviations from the chiral limit, and discuss possible realization in 2D materials.
\end{abstract}
\maketitle
\noindent\textit{Introduction.}--
Electronic band structures of 2D materials can be controlled and designed by manipulating superlattice structures. 
One well-known example is twisted bilayer graphene (TBG), where interference between the two layers makes the band structure angle dependent. Remarkably, 
at some ``magic angles'', isolated nearly flat topological bands arise~\cite{bistritzer2011moire,po2018fragile,song2019all}, which become exactly flat in the chiral limit~\cite{tarnopolsky2019origin}, and similar flat bands may arise in other twisted-bilayer systems as well, such as quadratic band crossing point (QBCP) bilayers~\cite{li2021magic,https://doi.org/10.48550/arxiv.2209.06524}.
This property of twisted bilayers makes them an exciting platform for studying strongly correlated phenomena such as unconventional superconductivity and correlated insulators~\cite{cao2018correlated,cao2018unconventional,padhi2018doped,lu2019superconductors,yankowitz2019tuning,polshyn2019large,xie2019spectroscopic,kerelsky2019maximized,jiang2019charge,choi2019electronic,padhi2019pressure,cao2020strange,zondiner2020cascade,wong2020cascade,nuckolls2020strongly,liu2020tunable,regan2020mott,wang2020correlated,xie2020nature,wu2020collective,su2020current,stefanidis2020excitonic,bultinck2020mechanism,padhi2021generalized,he2021symmetry}. 
In single layer moir\'e systems, exciting progress towards similar interference has been achieved via 
spatially varying electrostatic field, magnetic field, and elastic strain
field~\cite{jiang2017visualizing,milovanovic2020band,mao2020evidence,manesco2020correlations,manesco2021correlation,giambastiani2022electron,ghorashi2022topological,dong2022dirac,PhysRevLett.128.176406,gao2022untwisting,de2022network,Wan_Sarkar_Sun_Lin_2023}. 
However, for these systems, it was recently argued~\cite{gao2022untwisting} that (without a constant background magnetic field) exact flat bands cannot be achieved even in the chiral limit.

In this Letter, we study single layer systems with QBCPs under periodic strains. In contrast to Dirac points in graphene, where strain serves as a vector potential 
via minimal coupling $i\partial \rightarrow i\partial + A$~\cite{katsnelson2012graphene}, for QBCPs at time-reversal invariant momenta, such gauge-field-like couplings are prohibited by the time-reversal symmetry. Because $i\partial$ and strain fields have opposite parity under time reversal, the couplings allowed by symmetry for QBCPs take the form of $\partial\partial \rightarrow \partial\partial + A$, where $A$ is proportional to the strain field. In other words, the strain fields here provide a director potential with angular momentum $\ell=2$, instead of a vector with $\ell=1$. 

Remarkably, we find that this strain-field coupling induces exact flat bands, in strong analogy to TBG. Here, instead of controlling the twisting angle, we vary the strength of the strain field. In the chiral limit, exact flat bands are obtained as the strength of the strain field hits certain ``magic" values, and the flat bands carry Chern numbers $C=\pm 1$. Away from the chiral limit, these magic flat bands survive in a wide range of phase space, although their bandwidth is no longer exactly zero.
We prove analytically that these exact flat bands are protected by the same mathematical principles as magic flat bands in TBG, and thus in analogy to TBG, they are fragile topological bands and their quantum metric satisfy the trace condition. In addition, we find that their Berry curvature distributions are more uniform than the TBG flat bands, making them ideal for the realization of fractional Chern insulators ~\cite{wu2012,roy2014band,ledwith2020fractional,ledwith2021strong,wang2021exact,mera2021engineering,ledwith2022family}.

Despite these similarities, it is also worth highlighting that these strain-induced flat bands are due to a very different physical mechanism, and thus they exhibit some unique physical properties, sharply distinct from TBG. For example, these strain-induced flat bands can arise at the $\Gamma$ valley and only need a single layer, while twisted-bilayer magic flat bands require a finite wave vector away from $\Gamma$~\cite{angeli2021gamma} and interference between two layers. Secondly, twisted bilayer magic flat bands arise for various dispersion, e.g., Dirac and QBCP~\cite{li2021magic}, while the strain-induced magic flat bands can only emerge from QBCPs. 


\noindent\textit{QBCP, strain field and director potential.}--
Without requiring fine tuning, a stable QBCP can only arise at time-reversal ($T$) invariant momenta ($T\mathbf{k}=-\mathbf{k}$) with proper rotational symmetry~\cite{sun2009topological}, where $T$ and rotational symmetries protect the QBCP from being gapped out or splitting into Dirac points~\cite{nilsson2008electronic,sun2009topological, neto2009electronic}. Here we will focus on a QBCP with $T$ and 3- or 6-fold rotational symmetry, e.g., the $\Gamma$ point of a kagome metal. Up to a change of basis, the Hamiltonian near $\Gamma$ takes the following form
\begin{align}\label{eq:Hgamma}
H_{\Gamma}= - t_0 a^2\left(c_0  k^{2} \mathbbm{1}- (k_{x}^{2}-k_{y}^{2})\sigma_{x} - 2  k_{x}k_{y} \sigma_{y}\right),
\end{align}
where $\mathbbm{1}$, $\sigma_x$, $\sigma_y$ represent the $2\times 2$ identity and Pauli matrices respectively, and $k=|\mathbf{k}| $. Without loss of generality, we set $t_0=1$ and $a=1$. Here, $c_0$ determines the average effective mass of the two bands. At $c_0=0$, a chiral symmetry emerges, $\{H,\sigma_z\}=0$, and thus the dispersion of the two bands is symmetric around $E=0$. A nonzero $c_0$ breaks the chiral symmetry with dispersion similar to a kagome metal (Fig.~\ref{fig:ExactFlatBand}(b)). The special case of $c_0=\pm 1$ generates a flat band, as in the nearest-neighbor-hopping kagome lattice [Fig.~\ref{fig:ExactFlatBand}(a), also Supplemental material (SM)~\cite{SM2022}].

\begin{figure}[t]
     \centering
\includegraphics[scale=1]{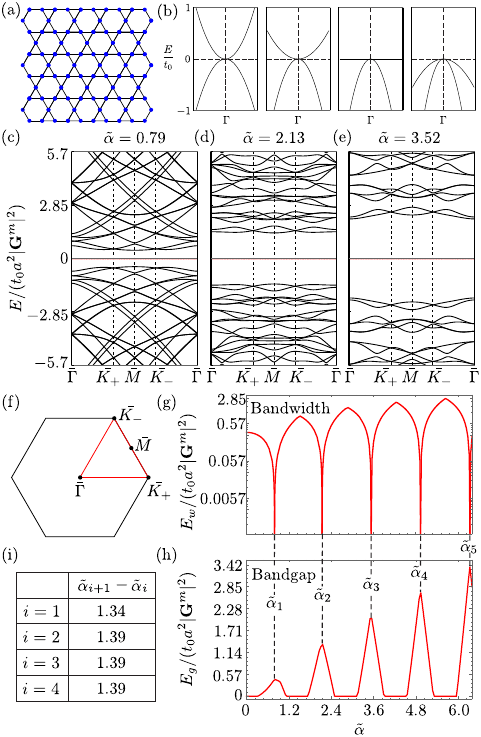}
     \caption{
     Exact flat bands from a QBCP under periodic strain.
     (a) Schematic of a kagome lattice, which has a QBCP at $\Gamma$.
     (b) The band structure of a QBCP [Eq.~\eqref{eq:Hgamma}] at $c_0=0,0.5,1,1.5$, where $c_0=0$ is the chiral limit.
     (c-e) Band structures in the chiral limit under strain $\tilde{A}$ [Eq.~\ref{eq:APhi} at $\phi = 0$] at different critical values of $\tilde{\alpha}=\alpha/(|\mathbf{G}^m| a)$ along the high symmetry path in the Moir\'e Brillouin zone shown in (f). There are two exact flat bands at $E=0$ in each case.
     (g,h) Bandwidth $E_w$ and bandgap $E_g$ as a function of $\tilde{\alpha}$, respectively. The bandwidth is exactly zero at the critical values of $\tilde{\alpha}$. The minima of the bandwidth occur at the same values of $\tilde{\alpha}$ as the maxima of the bandgap.
     (i) Table of the values of $\Delta \tilde{\alpha} = \tilde{\alpha}_{i+1}-\tilde{\alpha}_i$.
     }
     \label{fig:ExactFlatBand}
\end{figure}


Under a slowly varying strain field $u_{ij}(\mathbf{r})$, the Hamiltonian becomes
\begin{align}
     H(\mathbf{r})&=H_{\Gamma}+ A_{I}(\mathbf{r}) \mathbbm{1}+A_{x}(\mathbf{r}) \sigma_{x}+A_{y}(\mathbf{r}) \sigma_{y},\nonumber\\
     &=\left(
\begin{array}{cc}
 4 c_0\partial_z\partial_{\bar{z}}+A_I & 4\partial_z^2+\tilde{A}  \\
4\partial_{\bar{z}}^2+\tilde{A}^*& 4 c_0 \partial_z\partial_{\bar{z}}+A_I\\
\end{array}
\right),\label{eq:hamiltonian}
\end{align}
where $A_I\propto u_{xx}+u_{yy}$ describes bulk deformation, and $A_x\propto u_{xx}-u_{yy}$ and $A_y\propto u_{xy}$ describe shear deformations.
Here we introduce complex coordinates and strains
$z=x+iy$ and $\tilde{A} = A_x -i A_y$. Here, $\tilde{A}$ couples to $(k_x-i k_y)^2$, which is a director with angular momentum $\ell=2$. 
This director potential is the key reason why periodic strain can induce magic flat bands for QBCPs. In SM~\cite{SM2022}, we demonstrate a specific example using a kagome lattice tight-binding model to explicitly derive this Hamiltonian and the director potential coupling.

\noindent\textit{$C_{3v}$ symmetric and the chiral limit.}--
We start from the chiral limit $c_0=A_I=0$ and show that periodic strains lead to exact flat bands. A real physical system in general deviates from this ideal limit, but as is shown below, as long as such deviation is not too severe, magic flat bands still remain. Before studying this chiral limit, it is worth commenting on how to achieve this limit. To make $c_0$ close to zero, we need to use materials where the two bands near a QBCP have opposite effective mass (e.g., 
$\text{GaCu}_3(\text{OH})_6\text{Cl}_2$~\cite{mazin2014theoretical}). To make $A_I<<A_x$ or $A_y$, it simply requires to use 2D materials with a bulk modulus ($B$) larger than the shear modulus ($G$), so that shear deformations ($A_x$ and $A_y$) have lower energy cost than bulk deformation ($A_I$). In typical materials, this condition $B>G$ is naturally satisfied, unless the material is auxetic, i.e., has negative Poisson ratio.

In the chiral limit, we apply a periodic strain with $C_{3v}$ symmetry. Such strain fields can be categorized into two classes depending on whether the mirror plane 
is parallel or perpendicular to the reciprocal lattice vector $\mathbf{G}^m$. The first category (parallel) includes space groups $p3m1$ and $p6mm$, while the latter (perpendicular) gives $p31m$. Here we will focus on the first category. Within the first harmonic approximation, the strain can be written as
\begin{align}\label{eq:APhi}
    \tilde{A}(\mathbf{r})=t_0 \frac{\alpha^2}{2}\sum_{n=1}^{3} \omega^{n-1} \;\;\cos\left(\mathbf{G}_n^m \cdot \mathbf{r} +\phi\right),
\end{align}
where $\alpha^2>0$ is the strength of the strain, $\phi$ is an arbitrary phase and $\omega=\exp(2\pi i/3)$. $\mathbf{G}^m_{1} = \frac{4\pi}{\sqrt{3}a^m}(0,1)$, $\mathbf{G}^m_{2,3} = \frac{4\pi}{\sqrt{3}a^m}(\mp\sqrt{3}/2,-1/2)$ are the reciprocal lattice vectors. 
The second category has similar exact flat bands (see SM~\cite{SM2022}) and its strain field is
\begin{align}
    \tilde{A}(\mathbf{r})=\pm t_0 \alpha^2\sum_{n=1}^{3} \omega^{n-1} \;\;\exp \left(i \mathbf{G}_n^m \cdot \mathbf{r} \right).
\end{align}

\begin{figure}
     \centering
\includegraphics[scale=1]{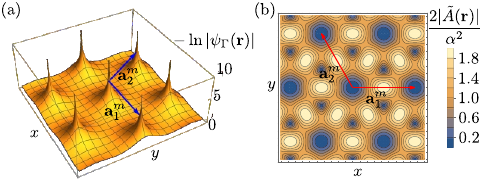}
     \caption{(a) The plot of $\ln(1/\vert \psi_\Gamma(\mathbf{r}) \vert)$ at the critical value $\tilde{\alpha} \approx 0.79$ for $\tilde{A}$ in Eq.~\eqref{eq:APhi} with $\phi = 0$. The spikes imply that $\psi_\Gamma(\mathbf{r})$ has zeros at $\mathbf{r}=n_1\mathbf{a}^{m}_1+n_2\mathbf{a}^{m}_2$, $n_1,n_2\in \mathbbm{Z}$. Here, $\mathbf{a}^{m}_1$ and $\mathbf{a}^{m}_2$ are the lattice vectors of superlattice. (b) Contour plot of $\frac{2\vert \tilde{A}(\mathbf{r}) \vert}{\alpha^2}$.} 
     \label{fig:WFabsA}
\end{figure}

\noindent\textit{Exact flat bands at $\phi=0$.}--We start by considering a special case with $\phi=0$ in Eq.~\eqref{eq:APhi}, where the system exhibits a higher rotational symmetry $C_{6v}$.
As shown in Fig.~\ref{fig:ExactFlatBand}(c-e), two exactly flat bands with $E=0$ arise at critical values of $\alpha$ (the square root of the strain strength). We define a dimensionless parameter $\tilde{\alpha}=\alpha/(|\mathbf{G}^m| a)$ ,which fully controls the physics of the system. These critical $\tilde{\alpha}$'s are roughly equally spaced $\Delta\tilde{\alpha}\approx 1.39$ [Fig.~\ref{fig:ExactFlatBand}(g, i)], and they also maximize the band gap that separates these two flat bands from others [Fig.~\ref{fig:ExactFlatBand}(h)].
In analogy to TBG in the chiral limit~\cite{tarnopolsky2019origin,gao2022untwisting}, these critical strains and exact flat bands can be analytically proved, and their Bloch wavefunctions can be analytically constructed. We start from the $\Gamma$ point. Because the strain preserves $T$ and $C_{3}$ symmetry, the two-fold degeneracy of the QBCP remains. In the chiral limit, the energy of these two degenerate states must remain at $E=0$, and their wavefunctions can be obtained from the Hamiltonian Eq.~\eqref{eq:hamiltonian}: $\Psi_{\Gamma,1}(\mathbf{r}) = \{\psi_\Gamma(\mathbf{r}),0\}$ and $\Psi_{\Gamma,2}(\mathbf{r}) = \{0,\psi_\Gamma^*(\mathbf{r})\}$, where $\psi_\Gamma(\mathbf{r})$ is a periodic function of the superlattice and obeys
$(4\partial_{\bar{z}}^2+\tilde{A}^*)\psi_\Gamma(\mathbf{r})=0$. $\Psi_{\Gamma,1}$ and $\Psi_{\Gamma,2}$ are related with each other via time-reversal, $T = \sigma_x K$, where $K$ is the complex conjugation.

If there are exact flat bands at $E=0$, the eigenfunctions can be written as $\{\psi_\mathbf{k}(\mathbf{r}),0\}$ and $\{0,\psi_{-\mathbf{k}}^*(\mathbf{r})\}$ with $(4\partial_{\bar{z}}^2+\tilde{A}^*) \psi_\mathbf{k}(\mathbf{r})=0$. Since the kinetic part of $D^\dagger(\mathbf{r})$ is antiholomorphic, we can construct a trial wave-function $\psi_\mathbf{k}(\mathbf{r})=f_\mathbf{k}(z)\psi_\Gamma(\mathbf{r})$. The $f_\mathbf{k}(z)$ needs to satisfy Bloch periodicity (translation by a superlattice vector $\mathbf{a}^m$ gives a phase shift of $e^{i \mathbf{k}\cdot \mathbf{a}^m}$). However, from Liouville's theorem, such $f_\mathbf{k}(z)$ must have poles, making $\psi_\mathbf{k}(\mathbf{r})$ divergent. 
To avoid such singularity, $\psi_\Gamma(\mathbf{r})$ needs to have a zero that cancels the pole of $f_\mathbf{k}(z)$. As shown in Fig.~\ref{fig:WFabsA}(a), at the magic values of $\tilde{\alpha}_i$, $\psi_\Gamma(\mathbf{0}) = 0$. Hence, we have
\begin{equation}\label{eq:WF}
\begin{split}
    \Psi_{\mathbf{k},1}(\mathbf{r}) &= \begin{bmatrix}\psi_{\mathbf{k}}(\mathbf{r}) \\0 \end{bmatrix}, \Psi_{\mathbf{k},2}(\mathbf{r}) = \begin{bmatrix}0\\\psi_{-\mathbf{k}}^*(\mathbf{r}) \end{bmatrix},\\
    \psi_\mathbf{k}(\mathbf{r})&=\frac{\vartheta_{\frac{\mathbf{k} \cdot \mathbf{a}^m_1}{2\pi}-\frac{1}{2},\frac{1}{2}-\frac{\mathbf{k} \cdot \mathbf{a}^m_2}{2\pi}}(\frac{z}{a_1},\omega)}{\vartheta_{-\frac{1}{2},\frac{1}{2}}(\frac{z}{a_1},\omega)}\psi_{\Gamma}(\mathbf{r}), 
\end{split}
\end{equation}
where $\vartheta_{a,b}(z,\tau)$ is the theta function of rational characteristic~\cite{TataI}, $\mathbf{a}_i^m$ are lattice vectors ($\mathbf{a}_1^m = a^m(1,0)$, $\mathbf{a}_2^m = a^m(-1,\sqrt{3})/2$) of the superlattice, $a_i = (\mathbf{a}_i^m)_x+ i (\mathbf{a}_i^m)_y$. As shown in the SM~\cite{SM2022}, 
these wavefunctions give two exact flat bands with $E=0$.  


In analogy to the Chern basis in TBGs~\cite{TBGIIBernevig}, because these two flat bands only involve holomorphic or antiholomorphic functions, they have Chern number $\pm 1$.
Same as Landau levels, their Fubini-Study metric $g(\mathbf{k})$ satisfies the trace condition $\text{tr}(g(\mathbf{k})) = |F_{xy}(\mathbf{k})|$~\cite{ledwith2021strong,ledwith2022family,roy2014band,wang2021exact,ledwith2020fractional,mera2021engineering}, where $F_{xy}(\mathbf{k})$ is the Berry curvature (see SM. Sec.~III~\cite{SM2022}). 
In SM Fig.~S1~\cite{SM2022}, the distributions of the Berry curvature in $k$-space are shown for the first three critical $\tilde{\alpha}$, which we found to be quite uniform. To quantify the non-uniformity of Berry curvature, we measure the ratio between the root-mean-square deviation of the Berry curvature and its average value
${\Delta F_{xy}}/\bar{F}_{xy}$~\cite{wu2012}.  The smallest value (most uniform distribution) is found at the second critical $\tilde{\alpha}$ with ${\Delta F_{xy}}/\bar{F}_{xy} \approx 0.027$, much smaller than the reported values in TBG flatbands~\cite{ledwith2020fractional}. Ideal quantum metric and very uniform Berry curvature make this system an ideal candidate for realizing fractional Chern insulators~\cite{wu2012,ledwith2021strong,ledwith2022family,roy2014band,wang2021exact,ledwith2020fractional,mera2021engineering}.



\begin{figure*}[t]
     \centering
\includegraphics[scale=1]{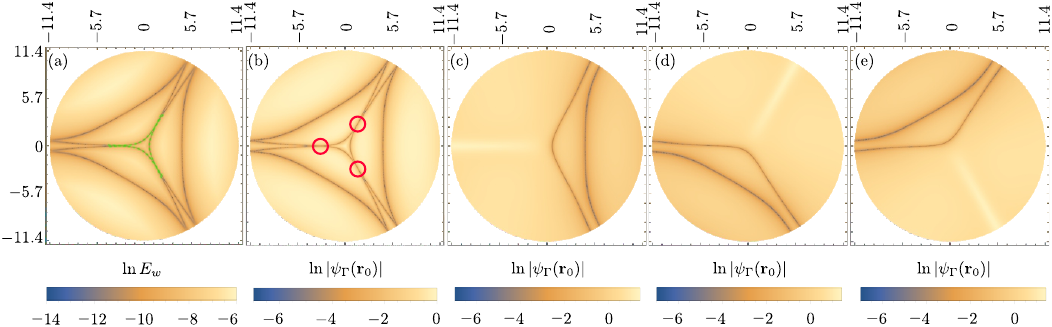}
    \caption{Exact flat bands in polar coordinate of $\tilde{\alpha}^2$ (radius) and $\phi$ (polar angle).
    (a) Bandwidth $\ln{E_w}$ as a function of $\phi$ and $\tilde{\alpha}$. The green dashed line shows the predicted value of principal critical $\tilde{\alpha}$ from $9$th order perturbation theory.
    (b) $\min\{\ln|\psi_\Gamma(\mathbf{r}=\mathbf{0})|,\ln| \psi_\Gamma(\mathbf{r}=(2\mathbf{a}_2^m+\mathbf{a}_1^m)/3)|, \ln|\psi_\Gamma(\mathbf{r}=(2\mathbf{a}_1^m+\mathbf{a}_2^m)/3)|\}$ as a function of $\tilde{\alpha}$ and $\phi$.
    (c-e) $\ln|\psi_\Gamma(\mathbf{r}=\mathbf{0})|$, $\ln| \psi_\Gamma(\mathbf{r}=(2\mathbf{a}_1^m+\mathbf{a}_2^m)/3)|$, $\ln|\psi_\Gamma(\mathbf{r}=(2\mathbf{a}_2^m+\mathbf{a}_1^m)/3)|$ as a function of $\tilde{\alpha}$ and $\phi$.
 The dark lines in each plot show critical $\tilde{\alpha}$ with two exact flat bands. At the crossing point of two dark lines [red circles in (b)], the strain induces four degenerate flat bands.}
     \label{fig:PhaseDiagram}
\end{figure*}

\noindent\textit{Generic $\phi$, double zeros and 4-fold flat bands.}-
In addition to $\phi=0$ (Eq.~\eqref{eq:APhi}, exact flat bands also arise at $\phi\ne 0$. 
At $\phi=n\pi/3$ ($\phi \ne n\pi/3$), the strain preserves 6-fold (3-fold) rotational symmetry with space group symmetry $p6mm$ ($p3m1$). 
In Fig.~\ref{fig:PhaseDiagram}(a), we plot the bandwidth as a function of $\tilde{\alpha}^2$ and $\phi$ in polar coordinate, setting $\tilde{\alpha}^2$ and $\phi$ to be the radius and polar angle, respectively. The dark lines mark the exact flat bands and the dashed green line marks analytic solution from perturbation theory (See SM~\cite{SM2022}). As we can see, for any value of $\phi$, exact flat bands can be reached at certain critical field strengths.

To further verify this conclusion, we show that $\psi_\Gamma(\mathbf{r})$ indeed contains zeros for all critical $\tilde{\alpha}$'s, and thus analytic wavefunctions can be constructed for these exact flat bands following the same procedure shown above. 
In Figs.~\ref{fig:PhaseDiagram}(c-e), 
we plot in log scale
$|\psi_\Gamma(\mathbf{r}=\mathbf{0})|$, $|\psi_\Gamma(\mathbf{r}=(2\mathbf{a}^m_1+\mathbf{a}^m_2)/3)|$ and $|\psi_\Gamma(\mathbf{r}=(2\mathbf{a}^m_2+\mathbf{a}^m_1)/3)|$, as well as $\min\{|\psi_\Gamma(\mathbf{r}=\mathbf{0})|, |\psi_\Gamma(\mathbf{r}=(2\mathbf{a}_2^m+\mathbf{a}_1^m)/3)|, |\psi_\Gamma(\mathbf{r}=(2\mathbf{a}_1^m+\mathbf{a}_2^m)/3)|\}$ in Fig.~\ref{fig:PhaseDiagram}(b). These three real space points are the 3-fold rotation centers of the $p3m1$ lattice, and the dark lines in these figures mark $\psi_\Gamma=0$, which perfectly match the exact flat bands shown in Fig.~\ref{fig:PhaseDiagram}(a).

The reason, we only see $\psi_\Gamma=0$ at these three high symmetry points in the unit cell, is that $\psi_\Gamma$ at other $\mathbf{r}$ are in general complex. To make both the real and imaginary parts zero, it typically requires to adjust two real control parameters simultaneously. In contrast, $\psi_\Gamma$ at these high symmetry points must be real, due to the $T$ and $C_{3v}$ symmetry (See SM~\cite{SM2022}), and thus its value can be tuned to zero with only one control parameter, which is why exact flat bands can emerge as we scan $\tilde{\alpha}$. This is why the addition of mirror symmetry to $C_3$ and $T$ is important.


It is easy to check that for these periodic strain fields, there exists a symmetry of $\phi\to \phi +2 \pi/3$. This transformation corresponds to a real space translation that swaps the three real-space high-symmetry points $\mathbf{0}$, $(2\mathbf{a}^m_1+\mathbf{a}^m_2)/3$, and  $(2\mathbf{a}^m_2+\mathbf{a}^m_1)/3$, as can be seen from Figs.~\ref{fig:PhaseDiagram}(c-e). This is the reason why Figs.~\ref{fig:PhaseDiagram}(a) and (b) are three-fold symmetric.
At points highlighted by the red circles in Fig.~\ref{fig:PhaseDiagram}(b), $\psi_\Gamma(\mathbf{r})$ has 2 zeros in a unit cell (two of the three real-space high-symmetry points). These double zeros are not accidental but due to the rotational symmetry. For example, at $\phi=\pi$, the rotational symmetry of the system increases to 6-fold and two of the high symmetry points, $(2\mathbf{a}^m_2+\mathbf{a}^m_1)/3$ and  $(2\mathbf{a}^m_1+\mathbf{a}^m_2)/3$, are connected by this 6-fold rotation. Thus, when $\psi_\Gamma(\mathbf{r}=(2\mathbf{a}_2^m+\mathbf{a}_1^m)/3)$ reaches zero at $\tilde{\alpha}\approx 1.695$, so does $\psi_\Gamma(\mathbf{r}=(2\mathbf{a}_1^m+\mathbf{a}_2^m)/3)$  (see SM. Fig.~S3(d)~\cite{SM2022} for the plot of $\psi_\Gamma(\mathbf{r})$). With two zeros, we can construct two meromorphic Bloch-periodic functions $f_\mathbf{k}^{(1)}(z)$ and $f_\mathbf{k}^{(2)}(z)$ similar to Eq.~\eqref{eq:WF}. Hence there are four flat bands instead of two, as is shown in SM. Fig.~S3~\cite{SM2022}. 
For completeness, we also plotted the predicted values of smallest critical $\tilde{\alpha}$ from $9$th order perturbation theory in Fig.~\ref{fig:PhaseDiagram}(a) (green dashed lines, see SM. Sec.~VII~\cite{SM2022} for details), they agree very well with the numerically calculated critical $\tilde{\alpha}$.

\noindent\textit{Fragile topology and exact solutions in interacting systems}--
As shown in SM Sec.~IX-X~\cite{SM2022}, in analogy to flat bands in TBG~\cite{po2018fragile,song2019all,TBGIIBernevig}, these flat bands are fragile topological bands. In addition, For spin-1/2 fermions with charge repulsion, these flat bands exhibit an emergent U(2)$\times$U(2) symmetry and integer fillings can be solved exactly (see SM Sec.~XII~\cite{SM2022}), similar to TBG~\cite{bultinck2020ground,lian2021twisted,hofmann2022fermionic}. At charge neutrality ($\nu=0$), there are degenerate ground states with Chern number $0$ or $\pm 2$, but thermal fluctuations stabilize the $C=0$ state via order-by-disorder. At filling $\nu=\pm 1$, the exact ground state is a Chern insulator with $C=\pm 1$.

\begin{figure}[b]
     \centering
\includegraphics[scale=1]{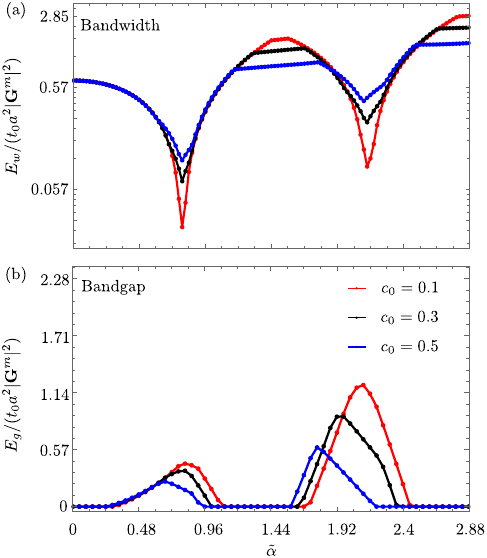}
     \caption{Effect of chiral symmetry breaking.(a)-(b)Bandwidth $E_w$ and bandgap $E_g$ as a function of $\tilde{\alpha}$ for different $c_0$.
     }
     \label{fig:C0Effect}
\end{figure}

\noindent\textit{Breaking chiral symmetry.}--
In analogy to TBG, as we move away from the chiral limit, these magic flat bands survive, although a small bandwidth emerges. Here we turn on the chiral symmetry breaking term $c_0$ and $A_I$ in Eq.~\eqref{eq:hamiltonian}, setting $\tilde{A}$ as Eq.~\eqref{eq:APhi} with $\phi=0$ and 
$A_I(\mathbf{r}) = - \frac{\alpha^2 c_0 t_0}{4}\sum_{i=1}^3\cos(\mathbf{G}_i^m\cdot\mathbf{r})$.
For a kagome lattice, this $A_I$ naturally arises from nearest-neighbor hoppings (See SM~\cite{SM2022}). As shown in Fig.~\ref{fig:C0Effect}, magic flat bands survive even if $c_0$ reaches $0.5$, although the bandwidth is no longer exactly zero, and the maximum bandgap and the minimum bandwidth slightly misalign. 
At $\tilde{\alpha}=0.79$, the bandgap to bandwidth ratio is $>17$
for $c_0=0.1$ and $>5$ for $c_0=0.3$.
Upon further increasing $c_0$, these two bands eventually mix with other bands. However, band hybridization and inversion result in an isolated highly flat band at $c_0=0.9$ and $\tilde{\alpha}=1.29$ as shown in SM. Fig.~S4~\cite{SM2022}. Similarly, for $\tilde{A}$ with $\phi=\pi$ [Eq.~\eqref{eq:APhi}] and the same $A_I$, at $c_0 \approx 0.15$ and $\tilde{\alpha} \approx 1.695$, the four degenerate flat bands around charge neutrality split into two pairs of isolated nearly flat bands as shown in 
SM. Sec.~IX~\cite{SM2022},
and both of them remain fragile topological, protected by $C_2$ and $C_3$, respectively.

\noindent\textit{Discussions.}--In this letter, we studied a QBCP near the $\Gamma$ point under periodic strain with $C_{3v}$ symmetry, and found exact topological flat bands in the chiral limit. 
Utilizing materials with nearly-chiral QBCPs (e.g., 
$\text{GaCu}_3(\text{OH})_6\text{Cl}_2$~\cite{mazin2014theoretical}) and a periodic strain field, which has already been achieved in experiments~\cite{jiang2017visualizing,mao2020evidence}, it is possible to access the vicinity of this chiral limit and explore these flat bands, 
offering a new platform to study topological flat bands, fragile topology, and correlated phases such as fractional Chern insulators and unconventional superconductivity. To further verify the feasibility of this proposal, in SM. Sec.~XIV~\cite{SM2022}, we compared our model with existing experiments and the estimation indicates current strain-engineering technology would allow us to reach at least the first two magic flat bands at the temperature $T\sim 4\ \mathrm{K}$. In addition, the same principle also applies to other systems with quadratic band crossings, such as photonic/phononic crystals, magnons, and optical lattices~\cite{sun2012topological,malki2019topological, li2020tuning, yang2022phononic, yu2022phononic1}.

\noindent \textit{Acknowledgements}.--We thank Rafael Fernandes for very helpful comments. This work was supported in part by the Office of Naval Research MURI N00014-20-1-2479 (XW, SS and KS) and Award N00014-21-1-2770 (XW and KS), and by the Gordon and Betty Moore Foundation Award N031710 (KS). The work at LANL (SZL) was carried out under the auspices of the U.S. DOE NNSA under contract No. 89233218CNA000001 through the LDRD Program, and was supported by the Center for Nonlinear Studies at LANL, and was performed, in part, at the Center for Integrated Nanotechnologies, an Office of Science User Facility operated for the U.S. DOE Office of Science, under user proposals $\#2018BU0010$ and $\#2018BU0083$.

\emph{Note added}: after the submission of this Letter, Ref.~\cite{eugenio2022twisted} showed that similar magic flat bands can arise in QBCP models with 4-fold rotational symmetry.

X.W. and S.S. contributed equally to this work.


\let\oldaddcontentsline\addcontentsline
\renewcommand{\addcontentsline}[3]{}
\bibliographystyle{apsrev4-1}
\bibliography{ref}
\pagebreak
\let\addcontentsline\oldaddcontentsline
\onecolumngrid

\makeatletter
\renewcommand \thesection{S-\@arabic\c@section}
\renewcommand\thetable{S\@arabic\c@table}
\renewcommand \thefigure{S\@arabic\c@figure}
\renewcommand \theequation{S\@arabic\c@equation}
\makeatother
\setcounter{equation}{0}  
\setcounter{figure}{0}  
\setcounter{section}{0}  

{
    \center \bf \large 
    Supplemental Material\vspace*{0.1cm}\\ 
    \vspace*{0.0cm}
}
\maketitle
\tableofcontents
\let\oldsection\section
\renewcommand\section{\clearpage\oldsection}
\section{Kagome Hamiltonian under strain}
Consider a buckled Kagome Hamiltonian,
\begin{equation}
H= \left(
\begin{array}{ccc}
 0 & -2t_{1} \cos{k_1} & -2t_{2} \cos{k_2}  \\
 -2t_{1} \cos{k_1} & 0 & -2t_{3} \cos{k_3}  \\
 -2t_{2} \cos{k_2}  & -2t_{3} \cos{k_3} & 0  \\
\end{array}
\right).
\end{equation}
Here $k_1=\mathbf{k} \cdot \mathbf{a}_1$,  $k_2=\mathbf{k} \cdot \mathbf{a}_2$, $k_3=\mathbf{k} \cdot \mathbf{a}_3$, $\mathbf{a}_1=a(1,0),\mathbf{a}_2=a(\frac{1}{2},\frac{\sqrt{3}}{2}),\mathbf{a}_3=\mathbf{a}_2-\mathbf{a}_1$.

To write the effective Hamiltonian near the quadratic band crossing point at $\Gamma$ of this Hamiltonian, we write the Hamiltonian in the eigenbasis at $\Gamma$:
\begin{equation}
H_{1}=U^{\dagger} H U \text{ where } U= \left(
\begin{array}{ccc}
 \frac{1}{\sqrt{3}} & \frac{-1}{\sqrt{2}} & \frac{-1}{\sqrt{6}}  \\
 \frac{1}{\sqrt{3}} & 0 & \frac{\sqrt{2}}{\sqrt{3}}  \\
 \frac{1}{\sqrt{3}} & \frac{1}{\sqrt{2}} & \frac{-1}{\sqrt{6}}  \\
\end{array}
\right).
\end{equation}
The new Hamiltonian $H_1$ has the form
\begin{equation}
H_{1}= \left(
\begin{array}{cc}
B & C  \\
C^{\dagger} & A \\
\end{array}
\right),
\end{equation}
where 
\begin{equation}
A= \left(
\begin{array}{cc}
2t_{2} \cos{k_2}  & \frac{1}{\sqrt{3}}(2t_{1} \cos{k1} -2t_{3} \cos{k_3})  \\
 \frac{1}{\sqrt{3}}(2t_{1} \cos{k1} -2t_{3} \cos{k_3})  & \frac{1}{3}(4t_{1} \cos{k_1}-2t_{2} \cos{k_2} +4t_{3} \cos{k_3} ) \\
\end{array}
\right),
\end{equation}
\begin{equation}
B=\frac{-4}{3}(t_{1} \cos{k}_1+t_{2} \cos{k}_2 +t_{3}\cos{k}_3 ),
\end{equation}
\begin{equation}
C=\left(
\begin{array}{cc}
\sqrt{\frac{2}{3}}(t_{1} \cos{k_1}-t_{3} \cos{k_3}) & -\sqrt{\frac{2}{3}}(t_{1} \cos{k_1}-2t_{2} \cos{k_2} +t_{3} \cos{k_3} ) \\
\end{array}
\right).
\end{equation}
Then we integrate out the non-degenerate band to write a effective $2\times2$ matrix for the two degenerate bands:
\begin{equation}
H'= A-C^{\dagger}B^{-1}C
\end{equation}
Before applying any strain field to the Kagome lattice, $t_{1}=t_{2}=t_{3}=t_0$, near the $\Gamma$ point, the effective $2\times2$ Hamiltonian is 
\begin{equation}
H_\Gamma= -\frac{1}{2}(-4+k_{x}^{2}+k_{y}^{2})t_0 a^2 \mathbbm{1}-k_{x}k_{y} t_0 a^2\sigma_{x}+\frac{1}{2}(k_{x}^{2}-k_{y}^{2})t_0 a^2\sigma_{z}.
\end{equation}
Here $t_0$ is the nearest-neighbor hopping strength, $a$ is the bond length.
When the strain field is nonzero, then $t_{1}\neq t_{2}\neq t_{3}$. Setting $t_{i}=t_0+\delta t_{i}$, and keeping only terms up to the lowest order in $\delta t_{i}$, the Hamiltonian around the $\Gamma$ point becomes (ignoring the constant part)
\begin{gather}
H(\mathbf{r}) = -\frac{1}{2}(k_{x}^{2}+k_{y}^{2})t_0 a^2 \mathbbm{1}+A_{I} \mathbbm{1}-k_{x}k_{y} t_0 a^2 \sigma_{x}+A_{x} \sigma_{x}+\frac{1}{2}(k_{x}^{2}-k_{y}^{2})t_0 a^2\sigma_{z}+A_{z}\sigma_{z},\\
A_{I}=\frac{2}{3}(\delta t_{1}+\delta t_{2}+\delta t_{3}),\\
A_{x}=\frac{2}{\sqrt{3}}(\delta t_{1}-\delta t_{3}),\\
A_{z}=-\frac{2}{3}(\delta t_{1}-2\delta t_{2}+\delta t_{3}).
\end{gather}
The above Hamiltonian can be mapped to the following one using the transformation $H\rightarrow \tilde{U}H\tilde{U}^\dagger$
where $\tilde{U} = e^{i\pi\sigma_y/4}e^{-i\pi\sigma_z/4}$:
\begin{gather}\label{EQ:13}
H_\Gamma= -\frac{1}{2}(k_{x}^{2}+k_{y}^{2})t_0 a^2 \mathbbm{1}+\frac{1}{2}(k_{x}^{2}-k_{y}^{2})t_0 a^2\sigma_{x}-k_{x}k_{y} t_0 a^2\sigma_{y},\\ \label{EQ:14}
H(\mathbf{r})=
-\frac{1}{2}(k_{x}^{2}+k_{y}^{2})t_0 a^2 \mathbbm{1} +A_{I} \mathbbm{1}-\frac{1}{2}(k_{x}^{2}-k_{y}^{2})t_0 a^2\sigma_{x}-A_{z}\sigma_{x} -k_{x}k_{y} t_0 a^2\sigma_{y}+A_{x} \sigma_{y}.
\end{gather}
In this new basis, the representation of time reversal operator becomes $T=\sigma_x K$.
To tune the coefficients of the terms proportional to $\mathbbm{1}$ in Eq.~\eqref{EQ:13}~\eqref{EQ:14}, we multiply them by $c_0$ to obtain (and multiply the whole matrix by $2$):
\begin{gather}
H_\Gamma= -c_{0}(k_{x}^{2}+k_{y}^{2})t_0 a^2 \mathbbm{1}-(k_{x}^{2}-k_{y}^{2}) t_0 a^2\sigma_{x}-2 k_{x}k_{y} t_0 a^2\sigma_{y},\\
H(\mathbf{r})= -c_{0}(k_{x}^{2}+k_{y}^{2})t_0 a^2 \mathbbm{1}+ 2c_{0} A_{I} \mathbbm{1}-(k_{x}^{2}-k_{y}^{2}) t_0 a^2\sigma_{x}-2A_{z} \sigma_{x}-2 k_{x}k_{y} t_0 a^2\sigma_{y}+2A_{x} \sigma_{y}.
\end{gather}
Now, if we set 
\begin{gather}
\delta t_1=-\frac{3}{16} t_0\alpha^{2} \cos(\mathbf{G}_{2}^m \cdot \mathbf{r}),\label{eq:delta_t1}\\
\delta t_2=-\frac{3}{16} t_0\alpha^{2} \cos(\mathbf{G}_{1}^m \cdot \mathbf{r}),\label{eq:delta_t2}\\
\delta t_3=-\frac{3}{16} t_0 \alpha^{2} \cos(\mathbf{G}_{3}^m \cdot \mathbf{r}),
\label{eq:delta_t3}
\end{gather}
we have
\begin{align}
A_{z}&=-\frac{1}{4} t_0 \alpha^{2} [\cos(\mathbf{G}_{1}^m \cdot \mathbf{r})-\frac{1}{2}\cos(\mathbf{G}_{2}^m \cdot \mathbf{r})-\frac{1}{2}\cos(\mathbf{G}_{3}^m \cdot \mathbf{r})],\\
A_{x}&=-\frac{1}{4} t_0 \alpha^{2} [\frac{\sqrt{3}}{2}\cos(\mathbf{G}_{2}^m \cdot \mathbf{r})-\frac{\sqrt{3}}{2}\cos(\mathbf{G}_{3}^m \cdot \mathbf{r})],\\
A_{I}&=-\frac{1}{8} t_0 \alpha^{2} [\cos(\mathbf{G}_{1}^m \cdot \mathbf{r})+\cos(\mathbf{G}_{2}^m \cdot \mathbf{r})+\cos(\mathbf{G}_{3}^m \cdot \mathbf{r})],
\end{align}
where $\mathbf{G}^m_{1} = \frac{4\pi}{\sqrt{3}a^m}(0,1)$, $\mathbf{G}^m_{2,3} = \frac{4\pi}{\sqrt{3}a^m}(\mp\sqrt{3}/2,-1/2)$ and $a^m$ is the lattice constant of the superlattice.

\section{Exact flat bands}

Since the continuum model in {\color{red}Eq.~(1)} of the main text is obtained by expanding around the quadratic band crossing point protected by $C_3$ and $T$, as long as the $\tilde{A}$ field does not break any of these symmetries, the lowest two bands at the $\Gamma$ point are degenerate with energy $E = 0$ if the chiral symmetry is not broken.  The two corresponding eigenfunctions are of the form $\Psi_{\Gamma 1} =\{\psi_\Gamma(\mathbf{r}),0\}^T$ and  $\Psi_{\Gamma 2} =\{0,\psi_\Gamma^*(\mathbf{r})\}^T$ satisfying 
\begin{equation}
(4\partial^2_{\bar{z}}+\tilde{A}^*)\psi_\Gamma(\mathbf{r}) = 0.
\end{equation}
To get an exact flat band for all wave vectors $\mathbf{k}$ with $E = 0$, we can construct a trial wave function
\begin{equation}\label{eq:psik}
\psi_\mathbf{k}(\mathbf{r})=f_\mathbf{k}(z) \psi_\Gamma(\mathbf{r})
\end{equation}
following~\citesupp{tarnopolsky2019origins}. Since $f_\mathbf{k}(z)$ is a holomorphic function, $\psi_\mathbf{k}(\mathbf{r})$ satisfies $(4\partial^2_{\bar{z}}+\tilde{A}^*)\psi_\mathbf{k}(\mathbf{r}) = 0$. However, for $\psi_\mathbf{k}(\mathbf{r})$ to be a wave function at the wave vector $\mathbf{k}$, it must satisfy the Bloch periodicity:
\begin{equation}
	\psi_\mathbf{k}(\mathbf{r}+\mathbf{a}^{m}_i) = e^{i \mathbf{k}\cdot \mathbf{a}^{m}_i} \psi_\mathbf{k}(\mathbf{r}),
\end{equation}
where $\mathbf{a}^{m}_i$ are the lattice vectors of the superlattice. Hence, $f_\mathbf{k}(z)$ must satisfy $f_\mathbf{k}(z+a_i) = e^{i \mathbf{k}\cdot\mathbf{a}^{m}_i}f_\mathbf{k}(z)$, where $a_i = (\mathbf{a}^{m}_i)_x+ i (\mathbf{a}^{m}_i)_y$, since $\psi_\Gamma(\mathbf{r})$ is a periodic function with $\psi_\Gamma(\mathbf{r}+\mathbf{a}^{m}_i) = \psi_\Gamma(\mathbf{r})$. But from Liouville's theorem, we know such  $f_\mathbf{k}(z)$ must have poles. This makes $\psi_\mathbf{k}(\mathbf{r})$ divergent, unless $\psi_\Gamma(\mathbf{r})$ has a zero exactly at the same position as the pole of $f_\mathbf{k}(z)$. We have shown in {\color{red}Fig.~2} of main text that at the critical value of $\alpha$, where the lowest two bands become exactly flat, $\psi_\Gamma(\mathbf{r})$ have zeros at the positions $\mathbf{r} = \mathbf{r}_0+ m\mathbf{a}^{m}_1+n\mathbf{a}^{m}_1 = (r_{01}+m)\mathbf{a}^{m}_1+ (r_{02}+n)\mathbf{a}^{m}_2$ ($m,n\in \mathbbm{Z}$). Then we can construct the following

\begin{equation}\label{eq:ftheta}
f_\mathbf{k}(z)=\frac{\vartheta_{\frac{\mathbf{k} \cdot \mathbf{a}^{m}_1}{2\pi}-\frac{1}{2}-r_{02},\frac{1}{2}-r_{01}-\frac{\mathbf{k} \cdot \mathbf{a}^{m}_2}{2\pi}}(\frac{z}{a_1},\omega)}{\vartheta_{-\frac{1}{2}-r_{02},\frac{1}{2}-r_{01}}(\frac{z}{a_1},\omega)}, 
\end{equation} 
where we choose $\mathbf{a}^{m}_1 = a^m(1,0)$, $\mathbf{a}^{m}_2 = a^m(-1/2,\sqrt{3}/2)$ (hence $a_1 = a^m, a_2 = a^m\omega$ where $\omega = -1/2+i\sqrt{3}/2$), and $\vartheta_{a,b}(z,\tau)$ is the theta function of rational characteristics $a$ and $b$ defined as~\citesupp{TataIs}:
\begin{equation}
	\vartheta_{a,b}(z,\tau) = \sum_{n=-\infty}^\infty e^{i \pi \tau (n+a)^2}e^{2\pi i(n+a)(z+b)}.
\end{equation}
Since $\vartheta_{a,b}(z,\tau)$ has zeros at $(n-1/2-a)\tau+(m+1/2-b)$ ($m,n\in \mathbbm{Z}$)~\citesupp{TataIs}, the zeros of $\vartheta_{-\frac{1}{2}-r_{02},\frac{1}{2}-r_{01}}(\frac{z}{a_1},\omega)$ cancel with the zeros of $\psi_\Gamma(\mathbf{r})$. Moreover, since
\begin{equation}
\begin{split}
\vartheta_{a,b}(z+1,\tau) &= \sum_{n=-\infty}^\infty e^{i \pi \tau (n+a)^2}e^{2\pi i(n+a)(z+1+b)}\\
 &= \sum_{n=-\infty}^\infty e^{2\pi i (n+a)}e^{i \pi \tau (n+a)^2}e^{2\pi i(n+a)(z+b)}\\
 &=  e^{2\pi i a}\sum_{n=-\infty}^\infty e^{i \pi \tau (n+a)^2}e^{2\pi i(n+a)(z+b)} =  e^{2\pi i a} \vartheta_{a,b}(z,\tau),
\end{split}
\end{equation}
(where we used $e^{2\pi i n} = 1$) and
\begin{equation}
\begin{split}
\vartheta_{a,b}(z+\tau,\tau) &= \sum_{n=-\infty}^\infty e^{i \pi \tau (n+a)^2}e^{2\pi i(n+a)(z+\tau+b)}\\
 &= \sum_{n=-\infty}^\infty e^{2\pi i \tau(n+a)}e^{i \pi \tau (n+a)^2}e^{2\pi i(n+a)(z+b)}\\
 &= e^{-i\pi \tau}\sum_{n=-\infty}^\infty e^{i \pi \tau (n+a+1)^2}e^{2\pi i(n+a)(z+b)}\\
 &= e^{-i\pi \tau}\sum_{n'=-\infty}^\infty e^{i \pi \tau (n'+a)^2}e^{2\pi i(n'-1+a)(z+b)}\\
 &= e^{-i\pi \tau-2\pi i (z+b)}\sum_{n'=-\infty}^\infty e^{i \pi \tau (n'+a)^2}e^{2\pi i(n'+a)(z+b)}\\
 &= e^{-i\pi \tau-2\pi i (z+b)}\sum_{n=-\infty}^\infty e^{i \pi \tau (n+a)^2}e^{2\pi i(n+a)(z+b)} = e^{-i\pi\tau-2\pi i (z+b)}  \vartheta_{a,b}(z,\tau),
\end{split}
\end{equation}
$f_\mathbf{k}(z)$ satisfies the Bloch periodicity. Therefore, 
\begin{equation}\label{EQ:WF}
\Psi_{\mathbf{k},1}(\mathbf{r}) = \begin{pmatrix}f_\mathbf{k}(z)\psi_\Gamma(\mathbf{r})\\0
\end{pmatrix} \text{ and } \Psi_{\mathbf{k},2}(\mathbf{r}) =\begin{pmatrix}0\\f_{-\mathbf{k}}^*(z)\psi_\Gamma^*(\mathbf{r})
\end{pmatrix}
\end{equation}
 are the eigenfunctions at any $\mathbf{k}$ with eigenvalues $E = 0$ as long as $\psi_\Gamma(\mathbf{r})$ has a zero in each superlattice unit cell. 

\section{Proof of \texorpdfstring{$\text{tr}(g) = |F_{xy}|$}{TEXT} for the wavefunction in {\color{red}Eq.~(5)} of the main text}
Here we show that for exactly flat bands with eigenfunctions of the form Eq.~\eqref{EQ:WF} ({\color{red}Eq.~(5)} of the main text) with Eq.~\eqref{eq:ftheta}, the trace of the Fubini-Study metric $\text{tr}(g(\mathbf{k}))$ is equal to the absolute value of the Berry curvature ($|F_{xy}(\mathbf{k})|$). To show this, we follow the arguments outlined in~\citesupp{ledwith2020fractionals}. The Fubini-Study metric $g_{ab}(\mathbf{k})$ and the Berry curvature $F_{xy}(\mathbf{k})$ are given by:
\begin{equation}
\begin{split}
g_{ab}(\mathbf{k}) &= \Re\left(\eta_{ab}(\mathbf{k})\right), F_{xy}(\mathbf{k}) = -2\Im\left(\eta_{xy}(\mathbf{k})\right),\\
\text{where }\eta_{ab}(\mathbf{k}) &= \frac{\langle \partial_a u_\mathbf{k}| \partial_b u_\mathbf{k}\rangle}{||u_\mathbf{k}||^2}-\frac{\langle \partial_a u_\mathbf{k}|u_\mathbf{k}\rangle\langle u_\mathbf{k}| \partial_b u_\mathbf{k}\rangle}{||u_\mathbf{k}||^4},
\end{split}
\end{equation}
where $u_\mathbf{k}(\mathbf{r})$ is the periodic part of the Bloch function, $\partial_a \equiv \partial_{k_a}$, and $||u_\mathbf{k}||^2 = \langle u_\mathbf{k} |u_\mathbf{k}\rangle$. 

Now, if $u_\mathbf{k}$ is holomorphic in $k$, i.e., $\partial_{\bar{k}} u_\mathbf{k} = \frac{1}{2}(\partial_{k_x} +i\partial_{k_y})u_\mathbf{k} = 0$, which we show below, then $\partial_{k_x} u_\mathbf{k} = (\partial_{k}+\partial_{\bar{k}}) u_\mathbf{k} = \partial_k u_\mathbf{k}$ and $\partial_{k_y} u_\mathbf{k} = i(\partial_{k}-\partial_{\bar{k}}) u_\mathbf{k} = i\partial_k u_\mathbf{k}$, and hence
\begin{equation}
\begin{split}
\eta_{xx}(\mathbf{k}) &= \frac{\langle \partial_k u_\mathbf{k}| \partial_k u_\mathbf{k}\rangle}{||u_\mathbf{k}||^2}-\frac{\langle \partial_k u_\mathbf{k}|u_\mathbf{k}\rangle\langle u_\mathbf{k}| \partial_k u_\mathbf{k}\rangle}{||u_\mathbf{k}||^4},\\
\eta_{xy}(\mathbf{k}) &= i \frac{\langle \partial_k u_\mathbf{k}| \partial_k u_\mathbf{k}\rangle}{||u_\mathbf{k}||^2}-i\frac{\langle \partial_k u_\mathbf{k}|u_\mathbf{k}\rangle\langle u_\mathbf{k}| \partial_k u_\mathbf{k}\rangle}{||u_\mathbf{k}||^4} = i\eta_{xx}(\mathbf{k}),\\
\eta_{yx}(\mathbf{k}) &= -i \frac{\langle \partial_k u_\mathbf{k}| \partial_k u_\mathbf{k}\rangle}{||u_\mathbf{k}||^2}+i\frac{\langle \partial_k u_\mathbf{k}|u_\mathbf{k}\rangle\langle u_\mathbf{k}| \partial_k u_\mathbf{k}\rangle}{||u_\mathbf{k}||^4} = -i\eta_{xx}(\mathbf{k}),\\
\eta_{yy}(\mathbf{k}) &= \frac{\langle \partial_k u_\mathbf{k}| \partial_k u_\mathbf{k}\rangle}{||u_\mathbf{k}||^2}-\frac{\langle \partial_k u_\mathbf{k}|u_\mathbf{k}\rangle\langle u_\mathbf{k}| \partial_k u_\mathbf{k}\rangle}{||u_\mathbf{k}||^4}= \eta_{xx}(\mathbf{k}).
\end{split}
\end{equation}
Hence we have,
\begin{equation}
\eta(\mathbf{k}) = \eta_{xx}(\mathbf{k})\begin{pmatrix}
1 & i\\
-i & 1
\end{pmatrix} \text{, and } g(\mathbf{k}) = \eta_{xx}(\mathbf{k})\begin{pmatrix}
1 & 0\\
0 & 1
\end{pmatrix}, F_{xy}(\mathbf{k}) = -2\eta_{xx}(\mathbf{k}),
\end{equation}
and consequently, $\text{tr}(g(\mathbf{k})) = |F_{xy}(\mathbf{k})|$. 

To show that $\partial_{\bar{k}} u_\mathbf{k} = 0$, where $u_\mathbf{k} = e^{-i\mathbf{k}\cdot \mathbf{r}}\psi_\mathbf{k}(\mathbf{r})$ is the periodic part of the Bloch function in Eq.~\eqref{eq:psik} with $f_\mathbf{k}(z)$ in Eq.~\eqref{eq:ftheta}, we first do a $\mathbf{k}$ dependent transformation $\tilde{\psi}_{\mathbf{k}}(\mathbf{r}) = \exp(-i \frac{a_1\bar{k}}{2}(\frac{1}{2}-r_{01}-\frac{k \bar{a}_2}{4\pi}-\frac{\bar{k} a_2}{8\pi})) \psi_{\mathbf{k}}(\mathbf{r}) = \exp(-i \frac{a_1\bar{k}}{2}(\frac{1}{2}-r_{01}-\frac{k \bar{a}_2}{4\pi}-\frac{\bar{k} a_2}{8\pi})) f_\mathbf{k}(z)\psi_{\Gamma}(\mathbf{r})$. (Note that this function $\tilde{\psi}_\mathbf{k}(\mathbf{r})$ is not periodic in $\mathbf{k}$-space, but that is not required for a well-defined Berry curvature~\citesupp{ledwith2020fractionals}.) Then, we have
\begin{equation}
\begin{split}
u_\mathbf{k}(\mathbf{r}) &= \exp(-i \frac{a_1\bar{k}}{2}(\frac{1}{2}-r_{01}-\frac{k \bar{a}_2}{4\pi}-\frac{\bar{k} a_2}{8\pi}))\exp(-i\mathbf{k}\cdot\mathbf{r}) f_\mathbf{k}(z)\psi_{\Gamma}(\mathbf{r})\\
 &= \exp(-i \frac{a_1\bar{k}}{2}(\frac{1}{2}-r_{01}-\frac{k \bar{a}_2}{4\pi}-\frac{\bar{k} a_2}{8\pi}))\exp(-i(k\bar{z}+\bar{k}z)/2) f_\mathbf{k}(z)\psi_{\Gamma}(\mathbf{r}).
\end{split}
\end{equation}
With this, we have
\begin{equation}
\begin{split}
&\partial_{\bar{k}} u_\mathbf{k}(\mathbf{r})\\
=&\partial_{\bar{k}} \left(\exp(-i \frac{a_1\bar{k}}{2}(\frac{1}{2}-r_{01}-\frac{k \bar{a}_2}{4\pi}-\frac{\bar{k} a_2}{8\pi}))\exp(-i(k\bar{z}+\bar{k}z)/2) f_\mathbf{k}(z)\psi_{\Gamma}(\mathbf{r})\right)\\
=&\frac{\psi_\Gamma(\mathbf{r})}{\vartheta_{-\frac{1}{2}-r_{02},\frac{1}{2}-r_{01}}(\frac{z}{a_1},\omega)} \partial_{\bar{k}} \left(\exp(-i \frac{a_1\bar{k}}{2}(\frac{1}{2}-r_{01}-\frac{k \bar{a}_2}{4\pi}-\frac{\bar{k} a_2}{8\pi}))\exp(-i(k\bar{z}+\bar{k}z)/2)\vartheta_{\frac{\mathbf{k} \cdot \mathbf{a}_1}{2\pi}-\frac{1}{2}-r_{02},\frac{1}{2}-r_{01}-\frac{\mathbf{k} \cdot \mathbf{a}_2}{2\pi}}(\frac{z}{a_1},\omega)\right)\\
=&\frac{a_1\psi_\Gamma(\mathbf{r})}{\vartheta_{-\frac{1}{2}-r_{02},\frac{1}{2}-r_{01}}(\frac{z}{a_1},\omega)} \partial_{a_1\bar{k}} \bigg(\exp(-i \frac{a_1\bar{k}}{2}(\frac{1}{2}-r_{01}-\frac{k \bar{a}_2}{4\pi}-\frac{\bar{k} a_2}{8\pi}))\exp(-i(k\bar{z}+\bar{k}z)/2)\\
&\phantom{\frac{\psi_\Gamma(\mathbf{r})}{\vartheta_{-\frac{1}{2}-r_{02},\frac{1}{2}-r_{01}}(\frac{z}{a_1},\omega)} \partial_{\bar{k}} (}\sum_{n=-\infty}^\infty e^{i \pi \omega (n+\frac{\mathbf{k} \cdot \mathbf{a}_1}{2\pi}-\frac{1}{2}-r_{02})^2}e^{2\pi i(n+\frac{\mathbf{k} \cdot \mathbf{a}_1}{2\pi}-\frac{1}{2}-r_{02})(z/a_1+\frac{1}{2}-r_{01}-\frac{\mathbf{k} \cdot \mathbf{a}_2}{2\pi})}\bigg)\\
=&\frac{a_1\psi_\Gamma(\mathbf{r})}{\vartheta_{-\frac{1}{2}-r_{02},\frac{1}{2}-r_{01}}(\frac{z}{a_1},\omega)}\exp(-i \frac{a_1\bar{k}}{2}(\frac{1}{2}-r_{01}-\frac{k \bar{a}_2}{4\pi}-\frac{\bar{k} a_2}{8\pi}))\exp(-i(k\bar{z}+\bar{k}z)/2)\\
&\phantom{\frac{\psi_\Gamma(\mathbf{r})}{\vartheta_{-\frac{1}{2}-r_{02},\frac{1}{2}-r_{01}}(\frac{z}{a_1},\omega)} (}\sum_{n=-\infty}^\infty  \bigg(-i \frac{1}{2}(\frac{1}{2}-r_{01}-\frac{k \bar{a}_2}{4\pi}-\frac{\bar{k} a_2}{4\pi})-i z/2a_1+2i\pi\omega(n+\frac{\mathbf{k} \cdot \mathbf{a}_1}{2\pi}-\frac{1}{2}-r_{02})\frac{1}{4\pi}\\
&\phantom{\frac{\psi_\Gamma(\mathbf{r})}{\vartheta_{-\frac{1}{2}-r_{02},\frac{1}{2}-r_{01}}(\frac{z}{a_1},\omega)} (\sum_{n=-\infty}^\infty\bigg(}-2\pi i (n+\frac{\mathbf{k} \cdot \mathbf{a}_1}{2\pi}-\frac{1}{2}-r_{02})\frac{\omega}{4\pi}+2\pi i \frac{1}{4\pi}(z/a_1+\frac{1}{2}-r_{01}-\frac{\mathbf{k} \cdot \mathbf{a}_2}{2\pi})\bigg)\times\\
&\phantom{\frac{\psi_\Gamma(\mathbf{r})}{\vartheta_{-\frac{1}{2}-r_{02},\frac{1}{2}-r_{01}}(\frac{z}{a_1},\omega)} (\sum_{n=-\infty}^\infty\bigg(-}e^{i \pi \omega (n+\frac{\mathbf{k} \cdot \mathbf{a}_1}{2\pi}-\frac{1}{2}-r_{02})^2}e^{2\pi i(n+\frac{\mathbf{k} \cdot \mathbf{a}_1}{2\pi}-\frac{1}{2}-r_{02})(z/a_1+\frac{1}{2}-r_{01}-\frac{\mathbf{k} \cdot \mathbf{a}_2}{2\pi})}\\
=0.
\end{split}
\end{equation}
This completes the proof.

\section{Berry curvature distribution}
\begin{figure}[h]
     \centering
\includegraphics[scale=1]{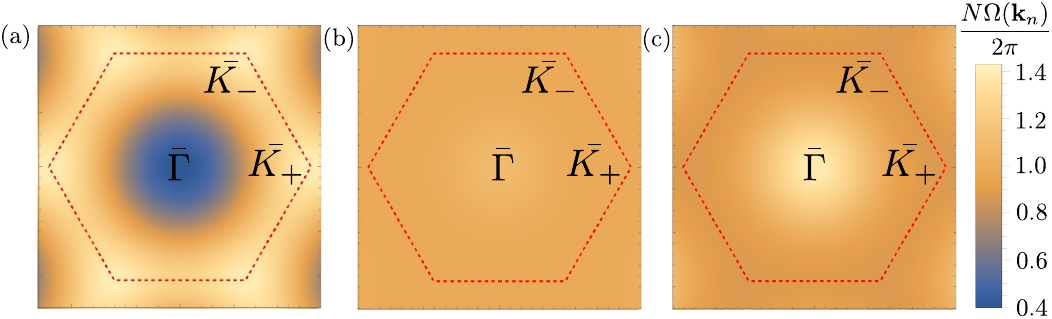}
     \caption{Berry Curvature distribution after adding $\sigma_z$ to separate the middle two bands at different critical values of $\tilde{\alpha}$ for the strain field defined in Eq.~(3) in the main text with $\phi=0$: (a) $\tilde{\alpha}=0.79$ (b) $\tilde{\alpha}=2.13$, and (c) $\tilde{\alpha}=3.52$. The calculated values of the measure $\Delta F_{xy}/\bar{F}_{xy}$ of the flucutuation of the Berry curvature defined in the main text for these three cases are $\Delta F_{xy}/\bar{F}_{xy} =0.287965$, $\Delta F_{xy}/\bar{F}_{xy} =0.0270411$, and $ \Delta F_{xy}/\bar{F}_{xy}=0.136598$, respectively. }
     \label{fig:berry curvature}
\end{figure}

\section{$\psi_{\Gamma}(\mathbf{r})$ is real at $C_{3v}$ invariant points in the superlattice unit cell}
We found earlier that if $\psi_\Gamma(\mathbf{r} = \mathbf{r}_0) = 0$, we can construct two exact flat bands with $E = 0$ and eigenfunctions are given in Eq.~\eqref{EQ:WF}. Generally, there is only one zero of $\psi_\Gamma(\mathbf{r})$ in one superlattice unit cell, otherwise we could construct more flat bands; this is why the zeros of $\psi_\Gamma(\mathbf{r})$ generally appear at isolated $C_3$ invariant points $\mathbf{r}_0$ in the superlattice unit cell. However, it is generally not guaranteed that at a $C_3$ invariant point $\psi_\Gamma(\mathbf{r})$ is real. Hence, if we want to set $\psi_\Gamma(\mathbf{r} = \mathbf{r}_0) = 0$ (where $\mathbf{r}_0$ is a $C_3$ invariant point) to obtain the critical value of the strain amplitude ($\alpha^2$ in the examples we show in the main text), we actually have two equations $\Re(\psi_\Gamma(\mathbf{r} = \mathbf{r}_0)) = 0 = \Im(\psi_\Gamma(\mathbf{r} = \mathbf{r}_0))$ and only one unknown $\alpha$ to tune, hence there may not be any solution for the critical $\alpha$. However, below we show that if there is extra mirror symmetry, $\psi_\Gamma(\mathbf{r})$ has to be real at the $C_{3v}$ symmetric points, hence there is only one equation $\Re(\psi_\Gamma(\mathbf{r} = \mathbf{r}_0)) = 0$ to satisfy to obtain the critical $\alpha$.

To show this we start from the $2\times 2$ Hamiltonian with chiral symmetry (S):
\begin{equation}
\mathcal{H}_0 = \int d^2r [c_1^\dagger(\mathbf{r})\,\, c_2^\dagger(\mathbf{r})] H(\mathbf{r})\begin{bmatrix}c_1(\mathbf{r})\\ c_2(\mathbf{r})\end{bmatrix},\,\,   H(\mathbf{r})= \begin{bmatrix}
 0 & 4\partial_z^2+\tilde{A}(\mathbf{r})  \\
4\partial_{\bar{z}}^2+\tilde{A}^*(\mathbf{r})& 0
\end{bmatrix},
\end{equation}
where $\tilde{A}$ is the strain field, $c_i^\dagger(\mathbf{r})$ are creation operators. In the following we prove that as long as the strain field $\tilde{A}$ has $C_3$, one mirror symmetry ($M_x$ or $M_y$, parallel or perpendicular to the reciprocal lattice vector $\mathbf{G}^m$, respectively) and time reversal symmetry $T$, $\psi_\Gamma(\mathbf{r} = \mathbf{r}_0)$ is real at the $C_{3v}$ invariant point $\mathbf{r}_0$. For simplicity, we use $M_y$ and $C_3$ to derive the result. Before adding any strain field, the matrix representations $\rho$ of $C_3$, $M_y$, $S$ and $T$ of the Hamiltonian are:
\begin{subequations}
\label{eq:Symmetries}
\begin{align}
\rho(C_3)&= 
\begin{pmatrix}
 \omega^2 & 0 \\
0 &  \omega \\
\end{pmatrix}, \rho(C_3) H(\mathbf{r}) \rho(C_3)^{-1} = H(C_3\mathbf{r}),\\
\rho(M_{x/y})&= \sigma_{x},\; \rho(M_{x/y}) H(\mathbf{r}) \rho(M_{x/y})^{-1} = H(M_{x/y}\mathbf{r}),\\
\rho(S)&= \sigma_{z},\; \rho(S) H(\mathbf{r}) \rho(S)^{-1} = -H(\mathbf{r}),\\
\rho(T)&= \sigma_{x},\; \rho(T) H^*(\mathbf{r}) \rho(T)^{-1} = H(\mathbf{r}),
\end{align}
\end{subequations}
where $\omega=-\frac{1}{2}+i\frac{\sqrt{3}}{2}$, $K$ is complex conjugate operator, and we have $R c_{i}^\dagger(\mathbf{r})R^{-1} = c_{j}^\dagger(R\, \mathbf{r}) [\rho(R)]_{ji}$ for $\forall R\in \{C_3,M_x,M_y,S,T\}$. We have the following (anti)commutation relations between the symmetries:
\begin{equation}
[C_3,S] = 0, [C_3,T] = 0, \{M_{x/y},S\} = 0, [M_{x/y},T] = 0, \{T,S\} = 0.
\end{equation}
After adding a strain field which has $M_y$ and $C_3$, the QBCP still remains at $\Gamma$ at $E = 0$ because the system still has $C_{3v}$, $S$ and $T$. Let the corresponding eigenbasis be $\tilde{c}_{\Gamma,1}^\dagger = \int d^2r [c_1^\dagger(\mathbf{r})\,\, c_2^\dagger(\mathbf{r})]\Psi_{\Gamma,1}(\mathbf{r})$ and $\tilde{c}_{\Gamma,2}^\dagger = \int d^2r[c_1^\dagger(\mathbf{r})\,\, c_2^\dagger(\mathbf{r})]\Psi_{\Gamma,2}(\mathbf{r})$ with $H(\mathbf{r})\Psi_{\Gamma,1}(\mathbf{r})=0$ and $H(\mathbf{r})\Psi_{\Gamma,2}(\mathbf{r})=0$. This eigenbasis transforms under the symmetries of the $\Gamma$ point in the following way:
\begin{subequations}
\begin{align}
C_3 \tilde{c}_{\Gamma,i}^\dagger C_3^{-1} &= \tilde{c}_{\Gamma,j}^\dagger [B_{C_3}]_{ji},\\
M_{x/y} \tilde{c}_{\Gamma,i}^\dagger M_{x/y}^{-1} &= \tilde{c}_{\Gamma,j}^\dagger [B_{M_{x/y}}]_{ji},\\
S \tilde{c}_{\Gamma,i}^\dagger S^{-1} &= \tilde{c}_{\Gamma,j}^\dagger [B_{S}]_{ji},\\
T \tilde{c}_{\Gamma,i}^\dagger T^{-1} &= \tilde{c}_{\Gamma,j}^\dagger [B_{T}]_{ji},
\end{align}
\end{subequations}
where the matrices $[B_{R}]$ are called the sewing matrices. This implies that $\Psi_{\Gamma,1}(\mathbf{r})$ and $\Psi_{\Gamma,2}(\mathbf{r})$ satisfy the following equations:
\begin{subequations}
\begin{align}
    \rho(C_3)\Psi_{\Gamma,i}(\mathbf{r}) &= \Psi_{\Gamma,j}(C_3\mathbf{r})[B_{C_3}]_{ji}\\
    \rho(M_{x/y})\Psi_{\Gamma,i}(\mathbf{r}) &= \Psi_{\Gamma,j}(M_{x/y}\mathbf{r})[B_{M_{x/y}}]_{ji}\\
    \rho(S)\Psi_{\Gamma,i}(\mathbf{r}) &= \Psi_{\Gamma,j}(\mathbf{r})[B_{S}]_{ji}\\
    \rho(T)\Psi_{\Gamma,i}^*(\mathbf{r}) &= \Psi_{\Gamma,j}(\mathbf{r})[B_{T}]_{ji}
\end{align}
\end{subequations}
In general,  wavefunctions have the form $\Psi_{\Gamma,1}(\mathbf{r}) = \{f_1(\mathbf{r}),g_1(\mathbf{r})\}$ and $\Psi_{\Gamma,2}(\mathbf{r}) = \{f_2(\mathbf{r}),g_2(\mathbf{r})\}$. However, we can gauge fix the sewing matrices in a certain way such that we have $\Psi_{\Gamma,1}(\mathbf{r}) = \{\psi_\Gamma(\mathbf{r}),0\}^T$ and $\Psi_{\Gamma,2}(\mathbf{r}) = \{0,\psi_\Gamma^*(\mathbf{r})\}^T$, and $\psi_\Gamma(\mathbf{r})$ is real at the $C_3$ symmetric points, which we choose to be the origin without loss of generality. To show this, we start by choosing $[B_{C_3}] = \text{Diag}\{\omega^2,\omega\}$ such that:
\begin{subequations}
\begin{align}
\begin{pmatrix}
\omega^2 & 0 \\
0 &  \omega \\
\end{pmatrix} \Psi_{\Gamma,1}(\mathbf{r}) &= \omega^2 \Psi_{\Gamma,1}(C_3\mathbf{r}) \Rightarrow f_1(\mathbf{r}) = f_1(C_3\mathbf{r}), \omega^2g_1(\mathbf{r}) = g_1(C_3\mathbf{r}),\\
\begin{pmatrix}
\omega^2 & 0 \\
0 &  \omega \\
\end{pmatrix} \Psi_{\Gamma,2}(\mathbf{r}) &= \omega \Psi_{\Gamma,2}(C_3\mathbf{r}) \Rightarrow \omega f_2(\mathbf{r}) = f_2(C_3\mathbf{r}), g_2(\mathbf{r}) = g_2(C_3\mathbf{r}).
\end{align}
\end{subequations}
This implies that at the $C_3$ symmetric point (the origin) $g_1(\mathbf{0}) = f_2(\mathbf{0}) = 0$. Moreover, since $C_3$ and $T$ commute with each other, we have 
\begin{equation}
C_3 T \tilde{c}_{\Gamma,1}^\dagger (C_3 T)^{-1} =T C_3 \Psi_{\Gamma,1}(\mathbf{r}) \tilde{c}_{\Gamma,1}^\dagger C_3^{-1} T^{-1}=T \omega^2 \tilde{c}_{\Gamma,1}^\dagger T^{-1}= \omega T\tilde{c}_{\Gamma,1}^\dagger T^{-1}. 
\end{equation}
This implies 
\begin{equation}
T\tilde{c}_{\Gamma,1}^\dagger T^{-1} = e^{i \theta}\tilde{c}_{\Gamma,2}^\dagger.
\end{equation}
We set $\theta=0$ to get $T\tilde{c}_{\Gamma,1}^\dagger T^{-1} = \tilde{c}_{\Gamma,2}^\dagger$. This implies
\begin{equation}
    \sigma_x \Psi_{\Gamma,1}^*(\mathbf{r}) = \Psi_{\Gamma,2}(\mathbf{r}) \Rightarrow f_2(\mathbf{r})=g_1^*(\mathbf{r}), g_2(\mathbf{r})=f_1^*(\mathbf{r}).
\end{equation}
Furthermore, since $C_3 M_{x/y} = M_{x/y} C_3^{-1}$, we require that $[B_{M_{x/y}}] = \sigma_x$ to obtain:
\begin{equation}
    \sigma_x \Psi_{\Gamma,1}(\mathbf{r}) = \Psi_{\Gamma,2}(M_{x/y}\mathbf{r}) \Rightarrow g_1(\mathbf{r}) = g_1^*(M_{x/y}\mathbf{r}), f_1(\mathbf{r}) = f_1^*(M_{x/y}\mathbf{r}).
\end{equation}
Lastly, since $[S,C_3] = 0$, we can fix $[B_S]=\sigma_z$ implying $g_1(\mathbf{r}) = 0$. Using all these and renaming $f_1$ as $\psi_\Gamma$, we get
\begin{equation}
\Psi_{\Gamma,1}(\mathbf{r})=
\begin{pmatrix}
\psi_\Gamma(\mathbf{r})\\
0\\
\end{pmatrix} \text{ and } \Psi_{\Gamma,2}(\mathbf{r})=
\begin{pmatrix}
0\\
\psi_\Gamma^*(\mathbf{r})\\
\end{pmatrix} \text{ with } \psi_\Gamma(\mathbf{r}) = \psi_\Gamma^*(M_{x/y}\mathbf{r})\text{ and }\psi_\Gamma(\mathbf{r}) = \psi_\Gamma(C_3\mathbf{r}).
\end{equation}
This means that on the mirror invariant lines (and hence on the $C_{3v}$ invariant points) in real space $\psi_\Gamma(\mathbf{r})$ is real as promised.

\section{General form of $C_{3v}$ symmetric strain field}
In this section, we show that the most general form of $\tilde{A}(\mathbf{r})$ that satisfies $C_3$ rotation symmetry and mirror symmetry $M_x$ and translation symmetry is of the form:
\begin{equation}
\tilde{A}(\mathbf{r})=\sum_{\mathbf{G}^m} a_{\mathbf{G}^m} [e^{i\phi}(e^{i \mathbf{G}^m \cdot \mathbf{r}}+\omega e^{i C_3 \mathbf{G}^m \cdot \mathbf{r}}+\omega^2 e^{i C_3^2 \mathbf{G}^m \cdot \mathbf{r}})+ e^{-i\phi}(e^{i M_y \mathbf{G}^m \cdot \mathbf{r}}+\omega e^{i C_3 M_y\mathbf{G}^m \cdot \mathbf{r}}+\omega^2 e^{i C_3^2 M_y \mathbf{G}^m \cdot \mathbf{r}})],
\label{eq:general A}
\end{equation}
where $\mathbf{G}^m$ is a reciprocal lattice vector of the superlattice, $a_{\mathbf{G}^m}$ is the amplitude at $\mathbf{G}^m$ and is real. Also notice that when we have mirror $M_x$, the amplitude of $\mathbf{G}^m$ is related to the amplitude at $M_y\mathbf{G}^m$. Any $\tilde{A}$ of this form has $C_{3v}$ symmetry, and consequently, the corresponding $\psi_\Gamma(\mathbf{r})$ is real at the $C_{3v}$ invariant points in the superlattice unit cell as shown in the previous section.

To this end, let us consider a strain field $\tilde{A}(\mathbf{r}) = \sum_{\mathbf{G}^m}\tilde{A}_{\mathbf{G}^m}e^{i\mathbf{G}^m\cdot\mathbf{r}}$, where $\tilde{A}_{\mathbf{G}^m}$ is the complex amplitude at the reciprocal lattice vector $\mathbf{G}^m$. We start with $C_3$. Since $C_3$ has representation $\rho(C_3) = \text{Diag}\{\omega^2, \omega\}$~\eqref{eq:Symmetries}, if $\tilde{A}$ obeys this symmetry, we have 
\begin{equation}
  \rho(C_3) H(\mathbf{r}) \rho(C_3)^{-1} = H(C_3\mathbf{r}) \Rightarrow \tilde{A}(C_3\mathbf{r}) = \omega \tilde{A}(\mathbf{r}).
\end{equation}
To satisfy this condition, if $\tilde{A}$ must have amplitudes $\tilde{A}_{C_3\mathbf{G}^m} = \omega \tilde{A}_{\mathbf{G}^m}$ and $\tilde{A}_{C_3^2\mathbf{G}^m} = \omega^2 \tilde{A}_{\mathbf{G}^m}$ at reciprocal lattice vectors $C_3\mathbf{G}^m$ and $C_3^2\mathbf{G}^m$, respectively.

Next, we consider the mirror symmetry $M_x$. Since $M_x$ has the representation $\rho(M_x) = \sigma_x$~\eqref{eq:Symmetries}, if $\tilde{A}$ obeys this symmetry, we have 
\begin{equation}
  \rho(M_x) H(\mathbf{r}) \rho(M_x)^{-1} = H(M_x\mathbf{r}) \Rightarrow \tilde{A}(M_x\mathbf{r}) = \tilde{A}^*(\mathbf{r}).
\end{equation}
Furthermore, we have the following:
\begin{equation}
    \tilde{A}^*(\mathbf{r}) = \sum_{\mathbf{G}^m}\tilde{A}_{\mathbf{G}^m}^*e^{-i\mathbf{G}^m\cdot\mathbf{r}},
\end{equation}
and
\begin{equation}
\begin{split}
    \tilde{A}(M_x\mathbf{r}) &= \sum_{\mathbf{G}^m} \tilde{A}_{\mathbf{G}^m} e^{i\mathbf{G}^m\cdot M_x\mathbf{r}}\\
    &= \sum_{\mathbf{G}^m} \tilde{A}_\mathbf{G} e^{i (M_x^{-1}\mathbf{G}^m)\cdot \mathbf{r}}\\
    &= \sum_{\mathbf{G}^m} \tilde{A}_\mathbf{G} e^{i (M_x\mathbf{G}^m)\cdot \mathbf{r}} \text{ since }M_x^{-1} = M_x\\
    &= \sum_{\tilde{\mathbf{G}}^m} \tilde{A}_{M_y\tilde{\mathbf{G}}^m} e^{i (M_xM_y\tilde{\mathbf{G}}^m)\cdot \mathbf{r}}\text{ with }\tilde{\mathbf{G}}^m = M_y\mathbf{G}^m \text{ and } \mathbf{G}^m = M_y^{-1}\tilde{\mathbf{G}}^m = M_y\tilde{\mathbf{G}}^m\\
    &= \sum_{\mathbf{G}^m} \tilde{A}_{M_y\mathbf{G}^m} e^{i (-\mathbf{G}^m)\cdot \mathbf{r}}\text{ where we used }M_x M_y \mathbf{G}^m = C_2 \mathbf{G}^m = -\mathbf{G}^m.
\end{split}
\end{equation}
Comparing the expressions for $\tilde{A}^*(\mathbf{r})$ and $\tilde{A}(M_x\mathbf{r})$, we find $\tilde{A}_{M_y\mathbf{G}^m} = \tilde{A}_{\mathbf{G}^m}^*$. The constraints on the amplitudes from $C_3$ and $M_x$ together give the general form of $\tilde{A}(\mathbf{r})$ written in Eq.~\eqref{eq:general A}.

\section{Analytical expression of $\psi_\Gamma(\mathbf{0})$ corresponding to {\color{red} Eq.~(3)} of the main text}
For the strain field introduced in {\color{red}Eq.~(3)} of the main text, assuming $\alpha$ to be a small parameter, one can solve for the zero energy eigenfunction $\psi_\Gamma(\mathbf{r})$ satisfying $D^\dagger(\mathbf{r})\psi_\Gamma(\mathbf{r}) = 0$ perturbatively, plugging the ansatz
\begin{equation}
    \psi_\Gamma(\mathbf{r}) = 1 +\alpha^2 u_1(\mathbf{r}) +\alpha^4 u_2(\mathbf{r})+ +\alpha^6 u_3(\mathbf{r})+\dots,
\end{equation}
to get the following equation at $n$th order in perturbation:
\begin{equation}
    4\alpha^2\partial_{\bar{z}}^2 u_n(\mathbf{r}) + \tilde{A}^* u_{n-1}(\mathbf{r})= 0
\end{equation}
with $u_0(\mathbf{r}) = 1$. Note that since $\psi_\Gamma(\mathbf{r})$ is a periodic function, $u_i(\mathbf{r})$ is also a periodic function. It can be easily checked that
\begin{equation}
    u_1(\mathbf{r}) = -\frac{\alpha^2}{2|\mathbf{G}_1^m| a^2}(\cos(\mathbf{G}_1^m \cdot \mathbf{r} +\phi)+ \cos(\mathbf{G}_2^m \cdot \mathbf{r}+\phi)+  \cos(\mathbf{G}_3^m \cdot \mathbf{r}+\phi)).
\end{equation}
The perturbation formula for $\psi_\Gamma(\mathbf{0})$ up to $9$th order is:
\begin{equation}
\begin{split}
    \psi_\Gamma(\mathbf{0}) =& 1 -\frac{1}{2} \left(3 \cos (\phi )\right)\left(\frac{\alpha}{|\mathbf{G}_1^m| a}\right)^2 +\left(\frac{1}{8}-\frac{9}{32} \cos (2 \phi )\right)\left(\frac{\alpha}{|\mathbf{G}_1^m| a}\right)^4+\left(\frac{17}{384} \cos (3 \phi )-\frac{393 \cos (\phi )}{6272}\right)\left(\frac{\alpha}{|\mathbf{G}_1^m| a}\right)^6\\
    &+\left(\frac{10741}{903168}-\frac{10741 \cos (2 \phi )}{2119936}-\frac{3041 \cos (4 \phi )}{401408}\right)\left(\frac{\alpha}{|\mathbf{G}_1^m| a}\right)^8\\
    &+\left(-\frac{63419689 \cos (\phi )}{36734251008}+\frac{2240465 \cos (3 \phi )}{814055424}-\frac{3970077 \cos (5 \phi )}{6783795200}\right)\left(\frac{\alpha}{|\mathbf{G}_1^m| a}\right)^{10}\\
    &+\left(-\frac{352813829131 \cos (2 \phi )}{1919976852684800}+\frac{152170241 \cos (6 \phi )}{976866508800}-\frac{872097414337 \cos (4 \phi )}{4706882029158400}+\frac{80468561821}{194397656334336}\right)\left(\frac{\alpha}{|\mathbf{G}_1^m| a}\right)^{12}\\
    &+\bigg(-\frac{5528291603578991773 \cos (\phi )}{272801393335850080665600}+\frac{37972267125602627 \cos (3 \phi )}{358687003655610040320}-\frac{8495976000041 \cos (7 \phi )}{451860674799206400}\\
    &\phantom{+\bigg(}-\frac{63936431554809217 \cos (5 \phi )}{8187805799496579809280}\bigg)\left(\frac{\alpha}{|\mathbf{G}_1^m| a}\right)^{14}\\
    &+\bigg(\frac{423059821189493221507 \cos (2 \phi )}{572573259558795826062950400}+\frac{2343534326486042059 \cos (6 \phi )}{315815366552010935500800}-\frac{4481219087893514849 \cos (8 \phi )}{3493463807785207385292800}\\
    &\phantom{+\bigg(}-\frac{4721900270905437779065457 \cos (4 \phi )}{2045886052583514457355216486400}+\frac{785315566262839677020797}{79666736498654970756936499200}\bigg)\left(\frac{\alpha}{|\mathbf{G}_1^m| a}\right)^{16}\\
    &+\bigg(\frac{172354469422917993236000320817767 \cos (\phi )}{175470658041537415365704103078647562240}+\frac{62291290190456796486924869 \cos (3 \phi )}{32512999970786662187158575513600}\\
    &\phantom{+\bigg(}+\frac{189653096110200866171238461 \cos (5 \phi )}{1785774373064771522040915768115200}+\frac{11440641141751693454081 \cos (9 \phi )}{30560821390504994206541414400}\\
    &\phantom{+\bigg(}-\frac{119966518304046968049670002623 \cos (7 \phi )}{465174475699948737845160475990425600}\bigg)\left(\frac{\alpha}{|\mathbf{G}_1^m| a}\right)^{18}+\mathcal{O}\left(\left(\frac{\alpha}{|\mathbf{G}_1^m| a}\right)^{20}\right).
\end{split}
\end{equation}
The expressions for $\psi_\Gamma((2\mathbf{a}^m_2+\mathbf{a}^m_1)/3)$ and $\psi_\Gamma((2\mathbf{a}^m_1+\mathbf{a}^m_2)/3)$ can be obtained by using the transformations $\phi \rightarrow\phi+2\pi/3$ and $\phi \rightarrow \phi+4\pi/3$, respectively.

\section{Example of exact flat band in a system where the strain field has $p31m$ symmetry}
\begin{figure}[h]
     \centering
\includegraphics[scale=1]{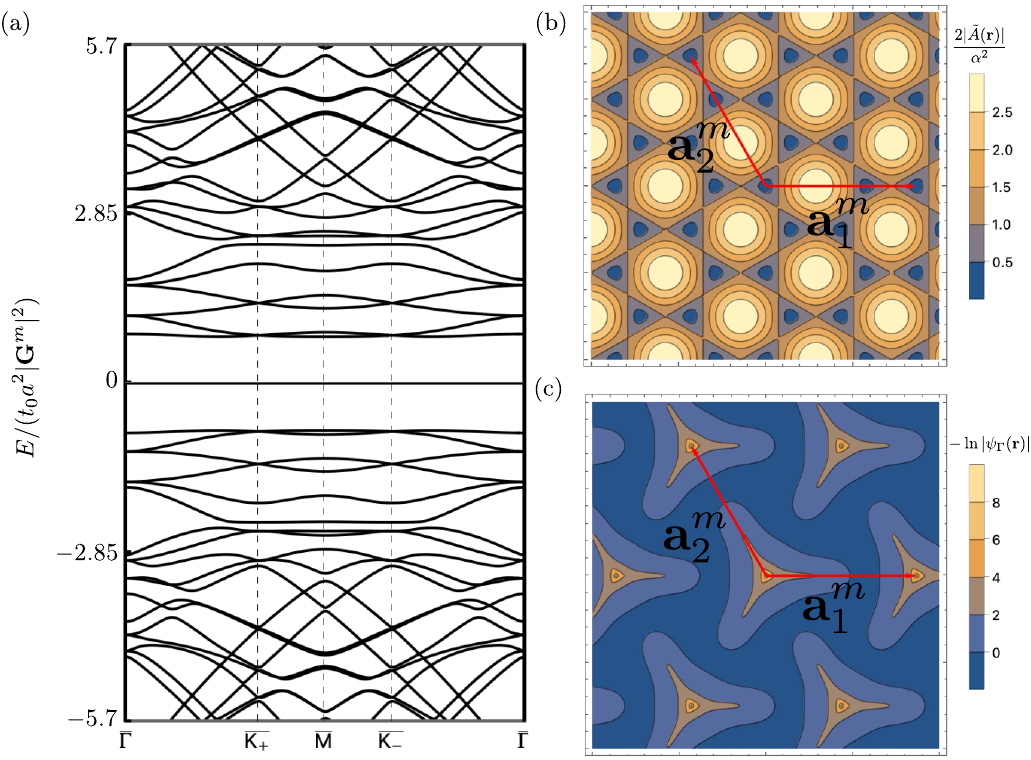}
     \caption{Exact flat band in chiral symmetric QBCP Hamiltonian with strain field $\tilde{A}= -\frac{\alpha^2}{2}(e^{i \mathbf{G}^m \cdot \mathbf{r}}+\omega e^{i C_3 \mathbf{G}^m \cdot \mathbf{r}}+\omega^2 e^{i C_3^2 \mathbf{G}^m \cdot \mathbf{r}})$. This strain field has $p31m$ space group symmetry.
     (a) Band structure corresponding to $\tilde{\alpha}=\alpha /(|\mathbf{G}^m| a) \sim 1.42$.
     (b) Plot of $\frac{2\vert \tilde{A}(\mathbf{r}) \vert}{\alpha^2}$ as a function of $\mathbf{r}$.
     (c) Plot $-\ln{\vert \psi_\Gamma(\mathbf{r}) \vert}$ as a function of $\mathbf{r}$ at $\tilde{\alpha} \sim 1.42$. The singularity at the origin indicates the zero of $\psi_\Gamma(\mathbf{0}$).
     }
     \label{fig:p31m}
\end{figure}

\section{Four flat bands at $\phi = \pi$ for the strain field in {\color{red} Eq.~(3)} of the main text}
\begin{figure}[h]
     \centering
\includegraphics[scale=0.9]{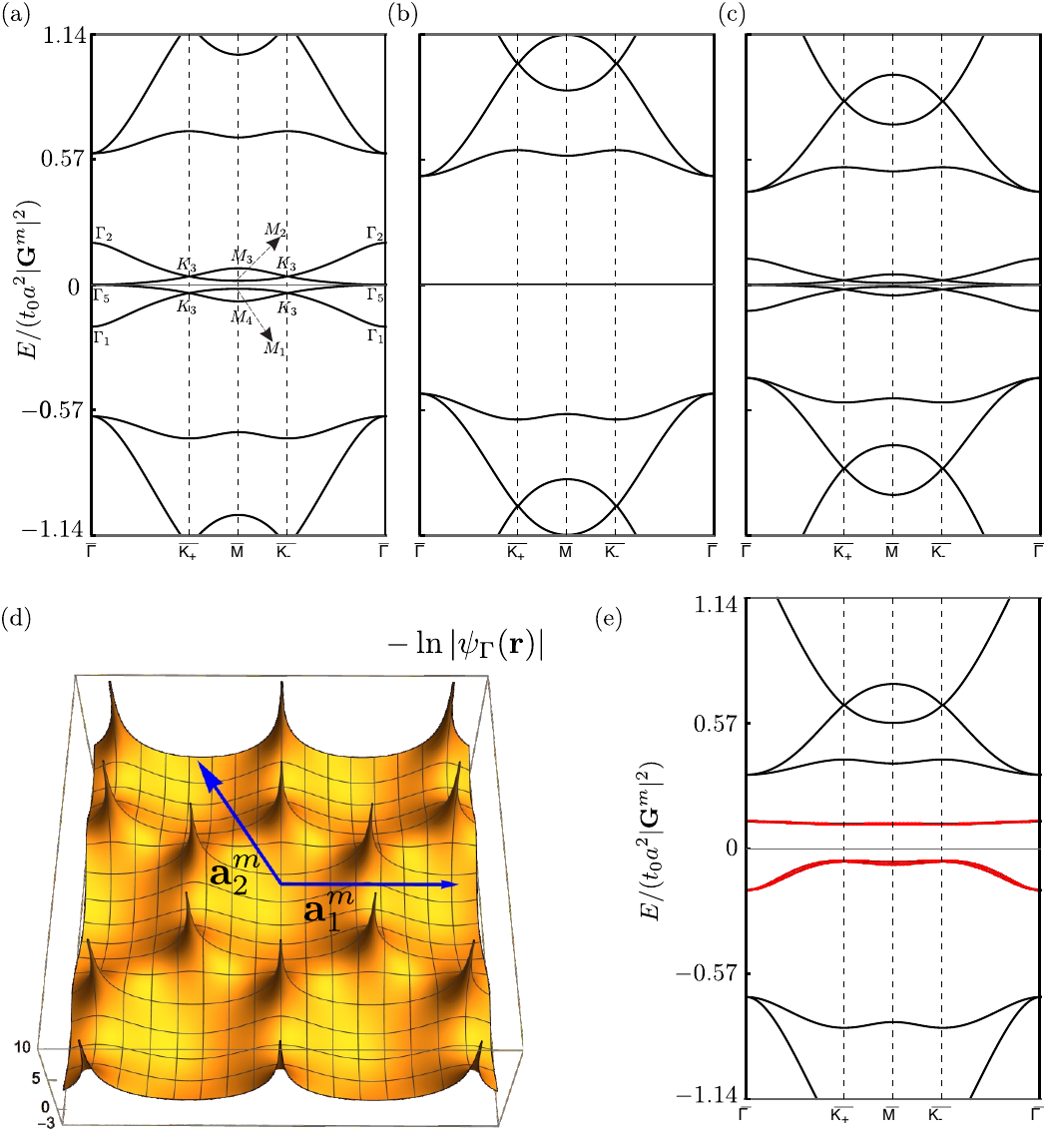}
     \caption{Four exact flat bands for the strain field in Eq.~(3) of the main text at $\phi =\pi$. (a) Band structure at $\tilde{\alpha}=1.43$.
     (b) Band structure at $\tilde{\alpha} \approx 1.695$.
     (c) Band structure at $\tilde{\alpha} =1.91$.
     (d) Plot of $-\ln{\vert \psi_\Gamma(\mathbf{r}) \vert}$ at $\tilde{\alpha} \approx 1.695$.
     (e) Band structure at $\tilde{\alpha} \approx 1.695$, and $c_0=0.15$. The set of four flat bands decomposes into two sets of nearly flat bands.
     }
     \label{fig:4flat band}
\end{figure}

\section{Fragile topology of the lowest two bands}
Here we show that in the chiral limit ($c_0 = 0$), the two lowest energy bands obtained from the continuum model in {\color{red}Eq.~(2)} of the main text are always fragile topological as long as: (i) the system retains three-fold rotation symmetry ($C_3$) with $[S,C_3] = 0$ and time reversal symmetry ($T$), (ii) the two bands are isolated from other bands. Since the continuum model is obtained from expanding around the QBCP protected by $C_3$ and $T$, so long as the $\tilde{A}$ field does not break any of these symmetries, the lowest two bands at the $\Gamma$ point are degenerate with energy $E = 0$ and have irrep labels $\Gamma_2\Gamma_3$, $\Gamma_3$, $\Gamma_3$ for space groups $p3$, $p3m1$, and $p31m$, respectively (here we use the notation of the Bilbao Crystallography Server (BCS)~\citesupp{aroyo2006bilbaoIs,aroyo2006bilbaoIIs,aroyo2011crystallographys}). Below we show that the two lowest bands are fragile topological for all four wallpaper groups ($p3$, $p3m1$, $p31m$ and $p6mm$) with $C_3$ symmetry.
\subsection{Space group \texorpdfstring{$p3$}{TEXT}}
In this case, the little co-group at the $M$ point is trivial, and the only irrep is $M_1$. At the $K$ point, there are three 1d irreps $K_1$, $K_2$ and $K_3$. Due to $[S,C_3]=0$, the two bands under consideration have the same irrep label at the $K$ point since the two bands are chiral symmetry partners of each other. Therefore, the only choices for the irrep labels at the high symmetry points (HSPs) are:
\begin{enumerate}
    \item $\Gamma_2\oplus\Gamma_3$ - $2M_1$ - $2K_1$,
    \item $\Gamma_2\oplus\Gamma_3$ - $2M_1$ - $2K_2$,
    \item $\Gamma_2\oplus\Gamma_3$ - $2M_1$ - $2K_3$.
\end{enumerate}
All these cases are fragile topological as listed in the SM of~\citesupp{song2020twisteds}. This is because none of these irrep labels can be obtained from symmetric exponentially localized Wannier functions (SLWF). In fact, these irrep labels are consistent with the following formal differences between SLWFs:
\begin{enumerate}
    \item $(^1E^2E\uparrow G)_{1b}\oplus (A_1\uparrow G)_{1a}\ominus (A_1\uparrow G)_{1c}$,
    \item $(^1E^2E\uparrow G)_{1c}\oplus (A_1\uparrow G)_{1b}\ominus (A_1\uparrow G)_{1a}$,
    \item $(^1E^2E\uparrow G)_{1a}\oplus (A_1\uparrow G)_{1c}\ominus (A_1\uparrow G)_{1b}$,
\end{enumerate}
where $a$, $b$ and $c$ are the maximal Wyckoff positions in the unit cell.

\subsection{Space group \texorpdfstring{$p3m1$}{TEXT}}
In this case, the little co-group at the $M_1$ point is $C_s$ with one mirror. But since the mirror symmetry $\sigma_x$ (Eq.~\eqref{eq:Symmetries}) anticommutes with the chiral symmetry $S =\sigma_z$ (Eq.~\eqref{eq:Symmetries}), if one of the two bands has irrep $M_1$ (even parity), the other one has $M_2$ (odd parity). At the $K$ point, the story is the same as in the case of the space group $p3$. Therefore, the only choices for the irrep labels at the HSPs are:
\begin{enumerate}
    \item $\Gamma_3$ - $M_1\oplus M_2$ - $2K_1$,
    \item $\Gamma_3$ - $M_1\oplus M_2$ - $2K_2$,
    \item $\Gamma_3$ - $M_1\oplus M_2$ - $2K_3$.
\end{enumerate}
All these cases are fragile topological as listed in the SM of~\citesupp{song2020twisteds}. This is because none of these irrep labels can be obtained from SLWFs. In fact, these irrep labels are consistent with the following formal differences between SLWFs:
\begin{enumerate}
    \item $(E\uparrow G)_{1b}\oplus (A_1\uparrow G)_{1a}\ominus (A_1\uparrow G)_{1c}$,
    \item $(E\uparrow G)_{1c}\oplus (A_1\uparrow G)_{1b}\ominus (A_1\uparrow G)_{1a}$,
    \item $(E\uparrow G)_{1a}\oplus (A_1\uparrow G)_{1c}\ominus (A_1\uparrow G)_{1b}$.
\end{enumerate}

\subsection{Space group \texorpdfstring{$p31m$}{TEXT}}
In this case, the irreps at the $M$ point are the same as in $p3m1$ since the mirror operator anticommutes with chiral symmetry. At the $K$ point, we know that the two bands have the same $C_3$ eigenvalues since $[C_3,S] = 0$, but different $M_y$ eigenvalues $\{M_y,S\} = 0$. This means that the irrep labels of the two bands at $K$ are $K_1$ and $K_2$. Therefore, the only choice for the irrep labels at the HSPs is:
\begin{enumerate}
    \item $\Gamma_3$ - $M_1\oplus M_2$ - $K_1\oplus K_2$.
\end{enumerate}
This is fragile topological as listed in the SM of~\citesupp{song2020twisteds}. This is because these irrep labels cannot be obtained from SLWFs. In fact, these irrep labels are consistent with the following formal difference between SLWFs:
\begin{enumerate}
    \item $(^1E^2E\uparrow G)_{2b}\ominus (E\uparrow G)_{1a}$.
\end{enumerate}
\subsection{Space group \texorpdfstring{$p6mm$}{TEXT}}
In this case, at the $\Gamma$ point, we have $C_{6v}$ symmetry. Before adding the periodic strain field, the 2d irrep at the $\Gamma$ point is $\Gamma_5$ since $\rho(C_2) = \mathbbm{1}$ (i.e., $H(\mathbf{r}) = H(C_2\mathbf{r})$ when $\tilde{A} = 0$) along with Eq.~\eqref{eq:Symmetries}. However, as we argue in the main text, as long as the strain field does not break these symmetries, the degeneracy remains at $E = 0$ at $\Gamma$ due to the commutation relation $[C_3,S] = 0$. Therefore, the irrep label is still $\Gamma_5$ at $\Gamma$ even when $\tilde{A} \neq 0$. At the $M$ point, which has little co-group $C_{2v}$, due to the anti-commutation between the mirrors and the chiral symmetry, the irrep labels can only be $M_1 \oplus M_2$ or $M_3\oplus M_4$. At the $K$ point, the story is the same as in the space group $p31m$. Therefore, the only choice for the irrep labels at the HSPs is:
\begin{enumerate}
    \item $\Gamma_5$ - $M_1\oplus M_2$ - $K_1\oplus K_2$,
    \item $\Gamma_5$ - $M_3\oplus M_4$ - $K_1\oplus K_2$.
\end{enumerate}
All these cases are fragile topological as listed in the SM of~\citesupp{song2020twisteds}. This is because none of these irrep labels can be obtained from SLWFs. In fact, these irrep labels are consistent with the following formal differences between SLWFs:
\begin{enumerate}
    \item $(E\uparrow G)_{2b}\ominus (E_1\uparrow G)_{1a}$,
    \item $(A_1\uparrow G)_{3c}\oplus (B_1\uparrow G)_{1a}\ominus (A_1\uparrow G)_{2b}$.
\end{enumerate}

\section{Fragile topology of the lowest four bands at \texorpdfstring{$\phi = (2n+1)\pi/3$}{TEXT} for a system under strain defined in {\color{red}Eq.~(3)} of main text}
At these values of $\phi$, the lowest $4$ bands are connected as is shown in Fig.~\ref{fig:4flat band}. Here we show that these $4$ bands are fragile topological. The irrep labels at the HSPs are $\Gamma_1 \oplus \Gamma_2 \oplus \Gamma_5 - M_1 \oplus M_2 \oplus M_3 \oplus M_4 - 2K_3$. These irreps cannot be obtained from SLWFs. These irreps correspond to the following:
\begin{equation}
   (A_2 \uparrow G)_{1a} \oplus (A_1 \uparrow G)_{2b}\oplus(A_1\uparrow G)_{3c} \ominus (A_1 \uparrow G)_{1a} \ominus (B_1 \uparrow G)_{1a}.
\end{equation}
These irrep labels of the $4$ bands can also be decomposed into two sets of fragile roots~\citesupp{song2020twisteds} $\Gamma_1 \oplus \Gamma_2 - M_1 \oplus M_2 - K_3$ and $\Gamma_5 - M_3 \oplus M_4 - K_3$ (or $\Gamma_1 \oplus \Gamma_2 - M_3 \oplus M_4 - K_3$ and $\Gamma_5 - M_1 \oplus M_2 - K_3$). The first set corresponds to fragile bands protected by $C_3$ symmetry, and the second set corresponds to fragile bands protected by $C_2$ symmetry.

\section{Emergent U$(2)\times$U$(2)$ symmetry and exact ground state of the interacting Hamiltonian in the chiral flat band limit}
Here we follow~\citesupp{bernevig2021twisteds,lian2021twisteds} to show the that Coulomb interacting Hamiltonian projected to the flat bands can be solved exactly. The Coulomb interaction term has the form
\begin{equation}
    \mathcal{H}_I = \frac{1}{2\Omega_\text{tot}} \sum_{\mathbf{q}\in \text{MBZ}}\sum_{\mathbf{G}} V(\mathbf{q}+\mathbf{G})\delta \rho_{-\mathbf{q}-\mathbf{G}}\delta \rho_{\mathbf{q}+\mathbf{G}},
\end{equation}
where $\Omega_\text{tot}$ is the area of the superlattice, $V(\mathbf{q}) = \frac{2\pi e^2}{\epsilon} \frac{\tan(\xi q/2)}{q}$ is the screened Coulomb potential, where $\xi$ is the screening length, $\mathbf{G}$ are the moir\'e reciprocal lattice vectors (we dropped the superscript $m$ here from the notation for moir\'e reciprocal lattice vectors for simplicity), MBZ is abbreviation of moir\'e Brillouin zone, and $\delta \rho_{\mathbf{q}+\mathbf{G}}$ is the density operator at the wave vector $\mathbf{q}+\mathbf{G}$ relative to the charge neutrality point:
\begin{equation}
    \delta \rho_{\mathbf{q}+\mathbf{G}} = \sum_{\mathbf{k}\in\text{MBZ}}\sum_{\mathbf{G}'}\sum_{\alpha,s} \left(c_{\mathbf{k}+\mathbf{q},\mathbf{G}+\mathbf{G}',\alpha,s}^\dagger c_{\mathbf{k},\mathbf{G},\alpha,s}-\frac{1}{2}\delta_{\mathbf{q},\mathbf{0}}\delta_{\mathbf{G},\mathbf{0}}\right),
\end{equation}
where $c_{\mathbf{k},\mathbf{G},\alpha,s}^\dagger$ is the $\alpha$th (of the basis in which the QBCP Hamiltonian is written) creation operator at the wave vector $\mathbf{k}+\mathbf{G}$ with spin $s$. In this notation, the non-interacting part of the Hamiltonian ($\mathcal{H}$ in Eq.~(S35)) has the form:
\begin{equation}
\begin{split}
    \mathcal{H}_0 &= \sum_{\mathbf{k}\in \text{MBZ}}\sum_{\mathbf{G},\mathbf{G}'} \sum_{\alpha,\beta,s} \left[h_{\mathbf{G},\mathbf{G}'}(\mathbf{k})\right]_{\alpha,\beta} c_{\mathbf{k},\mathbf{G},\alpha,s}^\dagger c_{\mathbf{k},\mathbf{G}',\beta,s},\\
    \left[h_{\mathbf{G},\mathbf{G}'}(\mathbf{k})\right] & = \begin{bmatrix} 0 & -(k_{-}+G_{-})^2 \delta_{\mathbf{G},\mathbf{G}'} + \tilde{A}(\mathbf{G} -\mathbf{G}')\\ -(k_{+}+G_{+})^2 \delta_{\mathbf{G},\mathbf{G}'} + \tilde{A}^*(\mathbf{G} -\mathbf{G}') & 0\end{bmatrix},
\end{split}
\end{equation}
where $k_{\pm} = k_x \pm i k_y$ and $G_{\pm} = G_x \pm i G_y$. To project the interacting part of the Hamiltonian to the lowest two bands, we first diagonalize the harmonic part:
\begin{equation}
    \sum_{\mathbf{G}',\beta}\left[h_{\mathbf{G},\mathbf{G}'}(\mathbf{k})\right]_{\alpha,\beta} u_{\mathbf{k},\mathbf{G}',\beta,n} = \epsilon_n(\mathbf{k})u_{\mathbf{k},\mathbf{G},\alpha,n},
\end{equation}
where $\{u_{\mathbf{k},n}\}$ is the eigenvector of the matrix $[h(\mathbf{k})]$ with eigenvalue $\epsilon_n(\mathbf{k})$. We denote the eigenvalues as $\dots \leq\epsilon_{-n}(\mathbf{k})\leq \dots \leq\epsilon_{-1}(\mathbf{k}) \leq\epsilon_{1}(\mathbf{k})\leq \dots \epsilon_{n}(\mathbf{k}) \leq \dots$, where $\epsilon_{-1}(\mathbf{k})$ and $\epsilon_{1}(\mathbf{k})$ are the eigenvalues with lowest magnitude. With this, we can define the energy band basis $c_{\mathbf{k},n,s}^\dagger = \sum_{\mathbf{G},\alpha}u_{\mathbf{k},\mathbf{G},\alpha,n} c_{\mathbf{k},\mathbf{G},\alpha,s}^\dagger$ (also, $c_{\mathbf{k},\mathbf{G},\alpha,s}^\dagger = \sum_n u_{\mathbf{k},\mathbf{G},\alpha,n}^*c_{\mathbf{k},n,s}^\dagger$). With this, we can rewrite the density operator
\begin{equation}
\begin{split}
    \delta \rho_{\mathbf{q}+\mathbf{G}} &= \sum_{\mathbf{k}\in\text{MBZ}}\sum_{\mathbf{G}'}\sum_{\alpha,s}\sum_{mn} \left( u_{\mathbf{k}+\mathbf{q},\mathbf{G}+\mathbf{G}',\alpha,m}^*u_{\mathbf{k},\mathbf{G}',\alpha,n}c_{\mathbf{k}+\mathbf{q},m,s}^\dagger c_{\mathbf{k},n,s}-\frac{1}{2}\delta_{\mathbf{q},\mathbf{0}}\delta_{\mathbf{G},\mathbf{0}}\right)\\
    &= \sum_{\mathbf{k}\in\text{MBZ}}\sum_{\mathbf{G}'}\sum_{\alpha,s}\sum_{mn}u_{\mathbf{k}+\mathbf{q},\mathbf{G}+\mathbf{G}',\alpha,m}^*u_{\mathbf{k},\mathbf{G}',\alpha,n} \left(c_{\mathbf{k}+\mathbf{q},m,s}^\dagger c_{\mathbf{k},n,s}-\frac{1}{2}\delta_{\mathbf{q},\mathbf{0}}\delta_{m,n}\right)\\
    &= \sum_{\mathbf{k}\in\text{MBZ}}\sum_{s}\sum_{mn}M_{m,n}(\mathbf{k},\mathbf{q}+\mathbf{G}) \left(c_{\mathbf{k}+\mathbf{q},m,s}^\dagger c_{\mathbf{k},n,s}-\frac{1}{2}\delta_{\mathbf{q},\mathbf{0}}\delta_{m,n}\right),
\end{split}
\end{equation}
where $M_{m,n}(\mathbf{k},\mathbf{q}+\mathbf{G}) = \sum_{\mathbf{G'},\alpha}u_{\mathbf{k}+\mathbf{q},\mathbf{G}+\mathbf{G}',\alpha,m}^*u_{\mathbf{k},\mathbf{G}',\alpha,n}$ is the form factor matrix. To project the interacting Hamiltonian, we just need to project the density operator to the lowest two bands:
\begin{equation}
\begin{split}
    \overline{\delta\rho}_{\mathbf{q}+\mathbf{G}} &= \sum_{\mathbf{k}\in\text{MBZ}}\sum_{s}\sum_{m,n=\pm1}M_{m,n}(\mathbf{k},\mathbf{q}+\mathbf{G}) \left(c_{\mathbf{k}+\mathbf{q},m,s}^\dagger c_{\mathbf{k},n,s}-\frac{1}{2}\delta_{\mathbf{q},\mathbf{0}}\delta_{m,n}\right),\\
    \mathcal{O}_{\mathbf{q},\mathbf{G}} &= \sqrt{V(\mathbf{q}+\mathbf{G})} \overline{\delta\rho}_{\mathbf{q}+\mathbf{G}},\\
    \overline{\mathcal{H}}_I &= \frac{1}{2\Omega_\text{tot}}\sum_{\mathbf{q}\in\text{MBZ}}\sum_{\mathbf{G}} V(\mathbf{q}+\mathbf{G}) \overline{\delta\rho}_{-\mathbf{q}-\mathbf{G}}\overline{\delta\rho}_{\mathbf{q}+\mathbf{G}}\\
    & = \frac{1}{2\Omega_\text{tot}}\sum_{\mathbf{q}\in\text{MBZ}}\sum_{\mathbf{G}} \mathcal{O}_{-\mathbf{q},-\mathbf{G}}\mathcal{O}_{\mathbf{q},\mathbf{G}}.
\end{split}
\end{equation}
Note that $\mathcal{O}(-\mathbf{q},-\mathbf{G}) = \mathcal{O}^\dagger(\mathbf{q},\mathbf{G})$ since $M_{mn}^*(\mathbf{k},\mathbf{q}+\mathbf{G}) = M_{nm}(\mathbf{k}+\mathbf{q},-\mathbf{q}-\mathbf{G})$. Hence, $\overline{\mathcal{H}}_I$ is positive semidefinite. The form factor matrix $M$ that goes into this projected interacting Hamiltonian is a $2\times 2$ matrix since we are considering only two bands. Therefore, it can be written in terms of Pauli matrices:
\begin{equation}
    M(\mathbf{k},\mathbf{q}+\mathbf{G}) =  \alpha_0(\mathbf{k},\mathbf{q}+\mathbf{G})\mathbbm{1}+ \alpha_1(\mathbf{k},\mathbf{q}+\mathbf{G})\zeta_1 + \alpha_2(\mathbf{k},\mathbf{q}+\mathbf{G})\zeta_2+ \alpha_3(\mathbf{k},\mathbf{q}+\mathbf{G})\zeta_3,
\end{equation}
where $\zeta_i$ are Pauli matrices, $\alpha_i(\mathbf{k},\mathbf{q}+\mathbf{G})$ are complex numbers. The symmetries of $\mathcal{H}_0$ constrain $\alpha_i(\mathbf{k},\mathbf{q}+\mathbf{G})$. Here we focus on 
chiral symmetry ($S$). 
This was already introduced in SM. Sec.~S-V. In the notation of this section, the representation matrix is:
\begin{equation}
\begin{split}
    S c_{\mathbf{k},\mathbf{G},\alpha,s}^\dagger S^{-1} &= \sum_{\mathbf{G}',\beta}c_{\mathbf{k},\mathbf{G}',\beta,s}^\dagger [\rho(S)]_{\mathbf{G}',\beta;\mathbf{G},\alpha},\\
    [\rho(S)]_{\mathbf{G}',\beta;\mathbf{G},\alpha} &= \delta_{\mathbf{G},\mathbf{G}'}(\sigma_z)_{\beta\alpha}.\\
\end{split}
\end{equation}
This implies that the eigenvectors satisfy $\{u_{\mathbf{k},n}\}$ the following:
\begin{equation}
\begin{split}
    [\rho(S)]\{u_{\mathbf{k},n}\} &= \sum_{m} \{u_{\mathbf{k},m}\} [B_S(\mathbf{k})]_{mn},
\end{split}
\end{equation}
where $[B_S(\mathbf{k})]$ is the sewing matrix. Specializing to the exact flat bands in the chiral limit, since we already know from Eq.~(5) of the main text that we can choose a gauge where the eigen wavefunctions for the exact flat bands are eigenfunctions of the chiral symmetry operator, we can gauge-fix $[B_S(\mathbf{k})]$ to be $[B_S(\mathbf{k})] = \zeta_3$. Then, the form factor matrix satisfies the following:
\begin{equation}
\begin{split}
    M_{m,n}(\mathbf{k},\mathbf{q}+\mathbf{G}) &= \sum_{\mathbf{G'},\alpha} u^*_{\mathbf{k},\mathbf{G}+\mathbf{G}',\alpha,m}u_{\mathbf{k},\mathbf{G}',\alpha,n}\\
    &= \sum_{\mathbf{G'},\alpha} \{[\rho(S)]\{u_{\mathbf{k},m}\}\}_{\mathbf{G}+\mathbf{G}',\alpha}^\dagger \{[\rho(S)]\{u_{\mathbf{k},n}\}\}_{\mathbf{G}',\alpha}\\
    &= \sum_{\mathbf{G'},\alpha} \sum_{m'n'} \{u_{\mathbf{k},m'}(\zeta_3)_{m'm}\}_{\mathbf{G}+\mathbf{G}',\alpha}^\dagger \{u_{\mathbf{k},n'}(\zeta_3)_{n'n}\}_{\mathbf{G}',\alpha}\\
    &= \sum_{\mathbf{G'},\alpha} \sum_{m'n'} (\zeta_3)_{m'm}^* u_{\mathbf{k},\mathbf{G}+\mathbf{G}',\alpha,m'} u_{\mathbf{k},\mathbf{G}',\alpha,n'}(\zeta_3)_{n'n}\\
    &= \sum_{m'n'} (\zeta_3)_{m'm}^* M_{m',n'}(\mathbf{k},\mathbf{q}+\mathbf{G}) (\zeta_3)_{n'n}\\
    \Rightarrow [M(\mathbf{k},\mathbf{q}+\mathbf{G})] &= [\zeta_3^\dagger M(\mathbf{k},\mathbf{q}+\mathbf{G}) \zeta_3].
\end{split}
\end{equation}
This means that $\alpha_1(\mathbf{k},\mathbf{q}+\mathbf{G}) = 0 = \alpha_2(\mathbf{k},\mathbf{q}+\mathbf{G})$, and $[M(\mathbf{k},\mathbf{q}+\mathbf{G})]$ is diagonal. Moreover, we know from Eq.~(5) of the main text that the corresponding eigen-wavefunctions have Chern numbers $\pm 1$. This means that for this gauge choice, the basis $\{c_{\mathbf{k},1,s}^\dagger,c_{\mathbf{k},-1,s}^\dagger\}$ is the Chern basis. Furthermore, since
\begin{equation}
\begin{split}
    S \mathcal{O}(\mathbf{q},\mathbf{G})S^{-1} &= \sum_{\mathbf{k},s} \sum_{m,n=\pm 1} \sqrt{V(\mathbf{q}+\mathbf{G})} M_{m,n}(\mathbf{k},\mathbf{q}+\mathbf{G}) \left(Sc_{\mathbf{k}+\mathbf{q},m,s}^\dagger c_{\mathbf{k},n,s}S^{-1} - \frac{1}{2}\delta_{\mathbf{q},\mathbf{0}}\delta_{m,n}\right)\\
    &= \sum_{\mathbf{k},s} \sum_{m,n=\pm 1} \sqrt{V(\mathbf{q}+\mathbf{G})} [\zeta_3^\dagger M(\mathbf{k},\mathbf{q}+\mathbf{G}) \zeta_3]_{mn} \left(c_{\mathbf{k}+\mathbf{q},m,s}^\dagger c_{\mathbf{k},n,s} - \frac{1}{2}\delta_{\mathbf{q},\mathbf{0}}\delta_{m,n}\right)\\
    &= \sum_{\mathbf{k},s} \sum_{m,n=\pm 1} \sqrt{V(\mathbf{q}+\mathbf{G})} M_{m,n}(\mathbf{k},\mathbf{q}+\mathbf{G}) \left(c_{\mathbf{k}+\mathbf{q},m,s}^\dagger c_{\mathbf{k},n,s} - \frac{1}{2}\delta_{\mathbf{q},\mathbf{0}}\delta_{m,n}\right)\\
    &= \mathcal{O}(\mathbf{q},\mathbf{G}),
\end{split}
\end{equation}
we have $S \overline{\mathcal{H}}_I S^{-1} = \overline{\mathcal{H}}_I$. However $\overline{\mathcal{H}}_I$ already has U$(2)$ spin rotation symmetry (the generators of the spin rotation symmetry are $\Sigma^a = \sum_\mathbf{k}\sum_{n=\pm 1} (s^a)_{s,s'} c_{\mathbf{k},n,s}^\dagger c_{\mathbf{k},n,s'}$, where $a = 0,1,2,3$, and $s^0$ is $2\times 2$ identity matrix, $s^1$, $s^2$ and $s^3$ are the Pauli matrices). Adding the chiral symmetry enlarges the symmetry group to U$(2)\times$U$(2)$ with generators 
\begin{equation}
\Sigma^{ab} = \sum_\mathbf{k}\sum_{m,n=\pm 1} (\zeta^a)_{m,n}(s^b)_{s,s'} c_{\mathbf{k},m,s}^\dagger c_{\mathbf{k},n,s'}, \text{ where }a\in\{0,3\},\text{ } b\in\{0,1,2,3\}.
\end{equation}

Furthermore, since $[M(\mathbf{k},\mathbf{q}+\mathbf{G})]$ is diagonal, with $M_{1,1}(\mathbf{k},\mathbf{q}+\mathbf{G}) = \alpha_0(\mathbf{k},\mathbf{q}+\mathbf{G})+\alpha_3(\mathbf{k},\mathbf{q}+\mathbf{G}) \equiv M_1(\mathbf{k},\mathbf{q}+\mathbf{G})$, $M_{1,-1}(\mathbf{k},\mathbf{q}+\mathbf{G}) = \alpha_0(\mathbf{k},\mathbf{q}+\mathbf{G})-\alpha_3(\mathbf{k},\mathbf{q}+\mathbf{G}) \equiv M_{-1}(\mathbf{k},\mathbf{q}+\mathbf{G})$ and $M_{1,-1} = M_{-1,1} = 0$, we can write the operator $\mathcal{O}_{\mathbf{q},\mathbf{G}}$ in the following way
\begin{equation}
    \mathcal{O}(\mathbf{q},\mathbf{G}) = \sum_{\mathbf{k},s} \sum_{e =\pm 1} \sqrt{V(\mathbf{q}+\mathbf{G})} M_{e}(\mathbf{k},\mathbf{q}+\mathbf{G}) \left(c_{\mathbf{k}+\mathbf{q},e,s}^\dagger c_{\mathbf{k},e,s} - \frac{1}{2}\delta_{\mathbf{q},\mathbf{0}}\right),
\end{equation}
where $e$ goes over the Chern band basis. Since the number of electrons $N$ is conserved, we can introduce a Lagrangian multiplier $A_\mathbf{G}$ to rewrite the projected interacting part of the Hamiltonian as:
\begin{equation}\label{eq:HIcp}
    \overline{\mathcal{H}}_I = \frac{1}{2\Omega_\text{tot}}\sum_\mathbf{G} \left[\left(\sum_\mathbf{q}\left(\mathcal{O}_{\mathbf{q},\mathbf{G}}-A_\mathbf{G} N \delta_{\mathbf{q},\mathbf{0}}\right)\left(\mathcal{O}_{-\mathbf{q},-\mathbf{G}}-A_{-\mathbf{G}} N \delta_{\mathbf{q},\mathbf{0}}\right)\right)+2A_{-\mathbf{G}}N\mathcal{O}_{\mathbf{0},\mathbf{G}} - A_{-\mathbf{G}}A_{\mathbf{G}}N^2\right],
\end{equation}
(see~\citesupp{lian2021twisteds} for details). At integer filling $\nu$ w.r.t. the charge neutrality point ($-2\leq \nu \leq 2$), we define spin polarized Fock state
\begin{equation}
    \ket{\Psi_\nu^{\nu^+,\nu^-}} = \prod_\mathbf{k} \prod_{j_1 =1}^{\nu^+} c_{\mathbf{k},+1,s_{j_1}}^\dagger \prod_{j_2 =1}^{\nu^-} c_{\mathbf{k},-1,s_{j_2}}^\dagger |0\rangle,
\end{equation}
where $\nu^+,\nu^- \in \{0,1,2\}$ with $\nu^++\nu^- = \nu+2$ and $s_{j_1}$, and $s_{j_2}$ can be chosen arbitrarily. Defining $A_\mathbf{G} = \sqrt{V(\mathbf{G})}\sum_{\mathbf{k}}[\frac{\nu^+-1}{N}M_1(\mathbf{k},\mathbf{G})+\frac{\nu^--1}{N}M_{-1}(\mathbf{k},\mathbf{G})]$, it can be shown that 
\begin{equation}
    \left(\mathcal{O}_{\mathbf{q},\mathbf{G}}-A_\mathbf{G} N \delta_{\mathbf{q},\mathbf{0}}\right) \ket{\Psi_\nu^{\nu^+,\nu^-}} = 0.
\end{equation}
At the flat metric limit $M_{m,n}(\mathbf{k},\mathbf{G}) = \xi(\mathbf{G})\delta_{m,n}$~\citesupp{lian2021twisteds} or at the charge neutrality point $\nu = 0$, the last two terms are constants; hence $\ket{\Psi_\nu^{\nu^+,\nu^-}}$ is a ground state. Away from the flat metric limit, the Fock state $\ket{\Psi_\nu^{\nu^+,\nu^-}}$ remains a ground state unless the gap between it and the excited states closes. In general, the flat metric condition is not satisfied except at $\mathbf{G} =\mathbf{0}$, but since the wavefunction decreases exponentially as $G$ increases~\citesupp{PhysRevB.103.205411s}, the flat metric condition is not largely violated.

Clearly, $\ket{\Psi_\nu^{\nu^+,\nu^-}}$ has the Chern number $(\nu^+-\nu^-)$. At the charge neutrality point $\nu = 0$, there are three choices: $(\nu^+ = 2, \nu^- = 0)$, $(\nu^+ = 0, \nu^- = 2)$  and $(\nu^+ = 1, \nu^- = 1)$. The first two cases have Chern numbers $C = \pm 2$, whereas the last case has Chern number $C = 0$. However, the last case is entropically more favorable since there are more ways of filling this than the other two. At other integer fillings, $\nu = \pm 1$, $\ket{\Psi_\nu^{\nu^+,\nu^-}}$ always carries a Chern number since in case of $\nu = - 1$, either $\nu^+$ or $\nu^-$ is zero, whereas in case of $\nu = + 1$, one of $\nu^+$ and $\nu^-$ is $1$, the other is $2$.

\section{Effect of breaking chiral symmetry}
\begin{figure}[h]
     \centering
\includegraphics[scale=1]{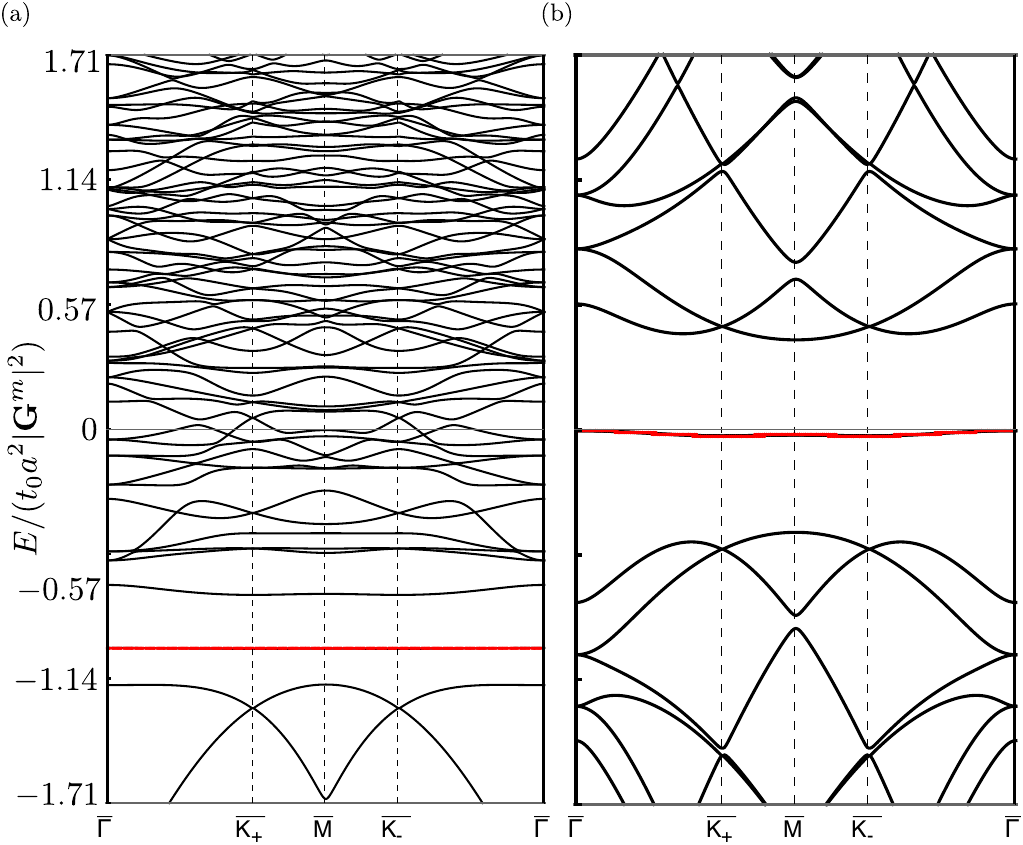}
     \caption{Effect of breaking chiral symmetry on the band structure. (a)Band Structure when $c_0=0.9$ (defined in {\color{red}Eq.~(2)}) and $\tilde{\alpha}=1.29$. The red band is the $n$th band among the total $2n$ bands.
     (b)Band structure when $c_0=0.1$ and $\tilde{\alpha}=0.79$, which is the first critical strength. The red bands are the $n$th and $(n+1)$th bands.
     }
     \label{fig:c0 sec}
\end{figure}

In Fig.~~\ref{fig:c0 sec}  we show two example band structures for $c_0 \neq 0$ in {\color{red}Eq.~(2)} of the main text along with the specific choice $A_I = - \frac{\alpha^2 c_0 t_0}{4}\sum_{i=1}^3\cos(\mathbf{G}_i^m\cdot\mathbf{r})$.

\section{Experimental feasibility}
In this section, we give an estimate of the experimentally achievable value of the nondimensional parameter $\alpha$ (defined in {\color{red}Eq.~(3)} in the main text) to argue that the critical strain strengths for exact flat bands can, in principle, be achieved in real experiments. The change in the hopping strength under applied strain in the long-wavelength limit can be quantified to be~\citesupp{suzuura2002phononss} 
\begin{equation}
\delta t_{i}=t_i -t_0=- \frac{\kappa \beta t_0}{a^2}\mathbf{a}_i \cdot (\mathbf{a}_i \cdot \nabla)\vec{u},   
\end{equation}
where $\kappa$ is the so-called reduction factor~\citesupp{suzuura2002phononss}, $\beta=-\frac{\partial \ln{t_0}}{\partial \ln{a}}$ and $\mathbf{a}_i$ is the vector corresponding to the bond. Plugging $\mathbf{a}_1=a(1,0),\,\mathbf{a}_2=a(\frac{1}{2}\frac{\sqrt{3}}{2})  \text{ and }\mathbf{a}_3=\mathbf{a}_2-\mathbf{a}_1$ for nearest neighbor kagome lattice, we get the following relations between strain tensor and modulation of hopping strength along different bond directions: 
\begin{equation}
\begin{split}
    \delta t_1&=- \kappa \beta t_0 u_{xx},\\
    \delta t_2&=- \kappa \beta t_0 (\frac{u_{xx}}{4}+\frac{3 u_{yy}}{4}+\frac{\sqrt{3}u_{xy}}{2}),\\
    \delta t_3&=- \kappa \beta t_0  (\frac{u_{xx}}{4}+\frac{3 u_{yy}}{4}-\frac{\sqrt{3}u_{xy}}{2}).
\end{split}
\end{equation}
To apply a strain field, we can either employ in-plane displacement field or out-of-plane displacement field, the latter is easier to achieve in experiments. Following \citesupp{mao2020evidences}, where a graphene monolayer was placed on a substrate to undergo a buckling transition, we consider periodic out-of-plane displacement field $h(\mathbf{r})$ to estimate experimentally accessible $\alpha$. For only out-of-plane displacements, we have
\begin{equation}
u_{ij}=\frac{1}{2} \partial_{i} h \partial_{j} h \propto {h_0}^{2} G^2 = {h_0}^{2} {(\frac{4\pi}{\sqrt{3}a^m})}^{2},
\end{equation}
where $h_0$ is the amplitude of out-of-plane displacement, $G$ is the length of the reciprocal lattice vectors of the superlattice, $a^m$ is the length of the lattice vectors of the superlattice. Hence, the magnitude of $\delta t_i$ is of order $\sim\kappa \beta t_0 {h_0}^{2} {(\frac{4\pi}{\sqrt{3}a^m})}^{2}$. 
We take $\kappa \approx \frac{1}{3}$ and $\beta \approx 2$ from~\citesupp{suzuura2002phononss}, then $\delta t_i \approx - \frac{2}{3} t_0 {h_0}^{2} {(\frac{4\pi}{\sqrt{3}a^m})}^{2}$. We use the values $h_0=0.17\text{ nm}$ and $a_m = (14.4 \pm 0.5)\text{ nm}$ from the experiments in~\citesupp{mao2020evidences} to get the estimate $\delta t_i \approx -0.0048908 t_0$. On the other hand, from Eq.~\eqref{eq:delta_t1}-\eqref{eq:delta_t3}, we know that the magnitude of $\delta t_i$ is $\frac{3}{16} t_0\alpha^{2}$. Therefore, $\delta t_i \approx -0.0048908 t_0$ corresponds to $\alpha \approx 0.161506$ and dimensionless $\tilde{\alpha} \approx 2.257$, which is larger than the first 2 critical $\tilde{\alpha}$ values shown in {\color{red}Fig.~1(c-e)} of the main text.

Now we estimate the temperature scale below which the upper bands are not important. Due to band folding in the Moir\'e system, the band gap is of the order $t_{0} (\frac{2\pi a}{\sqrt{3} a^{m}})^{2}$, where $t_{0}$ is the nearest-neighbor hopping strength, $a$ is the bond length and $a^m$ is the length of the lattice vectors of the superlattice. Here, we take the DFT calculation of $t_0 \approx 0.3 \text{eV}$ for $\text{GaCu}_3(\text{OH})_6\text{Cl}_2$(a material with nearly-chiral quadratic band crosing point) in~\citesupp{mazin2014theoreticals}).Using the $a^m = (14.4 \pm 0.5) \text{nm}$ and $a = \frac{0.246}{\sqrt{3}} \text{nm}$ from
the experiments in~\citesupp{mao2020evidences}, we get the the energy gap $\sim$1.54 meV, which corresponds to $T \approx 4.46\ \mathrm{K}$. In principle, as long as we keep the temperature below 4.46 K, thermal fluctuation will not destroy the flat band.
\bibliographystylesupp{apsrev4-1}
\bibliographysupp{ref.bib}

\end{document}